\documentclass[apj, twocolumn, twocolappendix, appendixfloats]{openjournal}
\usepackage{amsmath}
\usepackage{booktabs}
\usepackage{graphicx}
\usepackage[utf8]{inputenc}
\usepackage{hyperref}
\hypersetup{
    colorlinks=true,
    linkcolor=blue,
    filecolor=blue,
    urlcolor=blue,
    citecolor=blue,
}
\usepackage{orcidlink}
\usepackage{soul}
\usepackage{color, colortbl}
\usepackage{xcolor}
\usepackage[nolist]{acronym}
\usepackage{newtxtext, newtxmath, enumitem, multirow, ctable}
\usepackage[export]{adjustbox}
\usepackage{float}
\usepackage{placeins}

\usepackage{macros}

\graphicspath{{figure/}}

\begin{document}

\title{\vspace{-0.9 cm}
Hot or Cold? Radial Redistribution of Stars in FIRE Simulations of Milky Way-Mass Galaxies and the Asymmetry of Inward versus Outward Migrators
\vspace{-1.6 cm}}

\author{Cecilia Steel\,\orcidlink{0009-0009-9245-4651}$^1$}
\author{Andrew Wetzel\,\orcidlink{0000-0003-0603-8942}$^1$}
\author{Rori Kang$^{2,1}$}
\author{Fiona McCluskey$^1$}
\author{Sarah Loebman\,\orcidlink{0000-0003-3217-5967}$^3$}
\author{Kathryne J. Daniel\,\orcidlink{0000-0003-2594-8052}$^4$}

\affiliation{$^1$ Department of Physics \& Astronomy, University of California, Davis, CA 95616, USA}
\affiliation{$^2$ Department of Physics, Harvey Mudd College, Claremont, CA 91711, USA}
\affiliation{$^3$ Department of Physics, University of California, Merced, CA 95343, USA}
\affiliation{$^4$ Department of Astronomy \& Steward Observatory, University of Arizona, Tucson, AZ 85721, USA}

\begin{abstract}
Stars can radially redistribute (migrate) within galactic disks.
The degree to which this occurs as dynamically `cold' (preserves orbital eccentricity) or `hot' (increases eccentricity) remains debated.
Many models presume that radial redistribution occurs primarily via cold torquing, resulting in changes in angular momentum without dynamical heating.
We test the net dynamical heating associated with redistribution over stellar lifetimes using the FIRE cosmological zoom-in simulations of 12 Milky Way-mass galaxies.
We select star particles today that underwent significant changes in orbital angular momentum, $j_\phi$, since birth.
We investigate net changes in their orbital eccentricity, $e$, and we quantify the `cold-torqued' fraction of star particles with $|\Delta j_\phi/j_{\phi,\rm birth}|>0.2$ that preserved eccentricity ($|\Delta e|<0.1$) since birth.
The direction of radial redistribution is most critical: outward-migrating stars experienced smaller net changes in eccentricity, whereas inward-migrating stars almost always heat since birth.
For stars born on near-circular orbits ($e_{\rm birth}<0.2$), the cold-torqued fraction decreases rapidly with age today and is generally $<50\%$ at ages $\gtrsim2\Gyr$.
Stars born on moderately eccentric orbits ($e_{\rm birth}\approx0.4$) are the most likely to preserve their birth eccentricity.
However, the cold-torqued fraction is higher in earlier-forming and/or dynamically-colder disks.
Significantly, we identify a population of stars that dynamically `cooled', decreasing in eccentricity since birth: this is the primary way that stars end up on near-circular orbits today.
Overall, a star's migration direction, its $e_{\rm birth}$, and its age primarily determine whether it was dynamically heated, cooled, or unchanged.
In general, radial redistribution in FIRE is typically not cold between birth and today.
\end{abstract}

\section{Introduction}
\label{sec:Introduction}

Stars in a galaxy retain a record of how that galaxy formed over cosmic time, through their observed positions, velocities, elemental abundances, and ages. A key goal of much current research of the Milky Way (MW) is to reconstruct its history by using its stellar population as a record of its formation and evolution. Surveys such as Gaia \citep{Gaia2016}, SDSS-APOGEE \citep{Majewski2017}, LAMOST \citep{Cui2012}, GALAH \citep{DeSilva2015, Buder2018}, and SDSS-V Milky Way Mapper \citep{SDSS2025} now measure such properties for millions to billions of stars.
Furthermore, the upcoming Roman Galactic Plane Survey \citep{romangalacticplanesurvey} will extend such measurements up to $\gtrsim 20$ billion stars, reaching a significant fraction of all stars in the MW. However, linking the MW's present-day stellar populations to its past remains difficult, largely because of stellar `radial redistribution': stars currently may reside at galactocentric radii, $R$, far from where they formed. Consequently, stellar radial redistribution weakens the link between a star's present-day properties and its birth conditions, effectively scrambling the stellar record of a galaxy's assembly history. Many studies have tried to measure the extent of radial redistribution and to account for its effects on the spatial and velocity distributions of stars to recover their birth conditions and infer the formation and evolution of the MW \citep[for example][]{lu2024, Khoperskov2025, Ratcliffe2026, Dantas2025}.

Stellar atmospheres (approximately) preserve the elemental abundances of the gas out of which they formed, so stellar abundances provide a record of the state of the interstellar medium (ISM) when a star was born.
In disk galaxies, the metallicity of the ISM generally increases with time and declines with galactocentric radius (at least today), so in the absence of radial redistribution, one would expect relatively tight relations among stellar age, metallicity, and radius. Instead, observations in the solar neighborhood show a weak relation between stellar age and metallicity \citep{Edvardsson1993, Casagrande2011, Haywood2024} and include relatively old stars with super-solar metallicities \citep{Grenon1989, Castro1997}. By mixing stellar populations of different ages that formed at different radii, radial redistribution can blur the relations among age, metallicity, and galactocentric radius that galactic archeology relies on. In turn, radial redistribution can broaden the metallicity distribution \citep[for example][]{Grand2016a, Agertz2021} and relation between age and metallicity \citep{Casagrande2011, Haywood2024}, flatten metallicity radial gradients over time \citep[for example][]{Loebman2016, Graf2025a}, and complicate efforts to recover star-formation histories from present-day stellar populations \citep[for example][]{Minchev2025, Bernaldez2026}. Radial redistribution also may help shape the structure of the outer disk, including driving an upturn of mean stellar age \citep[for example][]{Roskar2008a, RadburnSmith2012, Herpich2017, Fiteni2026}, contributing to vertically flared distributions of coeval populations \citep[for example][]{Minchev2015}, and altering the azimuthal abundance distribution of stars \citep[for example][]{DiMatteo2013, Graf2025a}.

Because radial redistribution affects many of the observables we use to infer a galaxy's formation history, it is important to identify the signatures of the processes that drive it. One can think of two main modes of radial redistribution. The first is through dynamical scattering/heating, which results in a change in a star's random energy or eccentricity and can include a change in orbital angular momentum (or azimuthal action, $j_\phi$) and thus `guiding center radius' (the radius of a circular orbit with the same $j_\phi$). Over their lifetimes, stars undergo repeated interactions and perturbations within the disk, such as scattering off giant molecular clouds \citep[GMCs;][]{SpitzerSchwarzschild1951, Wielen1977, Lacey1984, Modak2026}, spiral arms \citep{BarbanisWoltjer1967, CarlbergSellwood1985, MinchevQuillen2006}, and bars \citep{Saha2010, Grand2016a}, and with perturbations external to the disk, such as satellites, merging galaxies \citep{Quillen2009, Bird2012, Carr2022}, or accretion of gas \citep{Bao2024}. 

If stars form on circular orbits, it should be straightforward to identify whether a star has undergone dynamical heating. However, that assumption is more appropriate for young stars born in relatively settled disks than for older stellar populations. Many works show that galaxy disks form vertically `upside down', such that older stars were born thicker and dynamically hotter, while younger populations formed progressively colder as the disk settled \citep[for example][]{Ma2017, Bird2021, Belokurov2022, Khoperskov2025}.
Cosmological simulations show that older stellar populations were born on hotter orbits than younger populations \citep[for example][]{verma2021, santucci2023, McCluskey2024}. Some of the present day kinematics of old stars may therefore reflect their birth conditions, rather than subsequent dynamical heating. Observations of galaxies at high redshifts indicate that MW-mass progenitor galaxies were not as thin or dynamically settled as the MW today \citep[for example][]{Kassin2012, Glazebrook2013, Conselice2014, Ubler2019, Wisnioski2019, Danhaive2025}, but instead were thicker, clumpier, and more turbulent, with higher gas fractions, burstier star formation, and more frequent mergers \citep[for example][]{MadauDickinson2014, Tacconi2020, FoersterSchreiberWuyts2020, jolly2026}. While many works have focused on the radial motions of stars, comparatively fewer works have focused on the radial motions of the gas from which they form. Because stars inherit the dynamics of that gas, older stars likely did not begin on circular orbits. This makes the signatures of radial redistribution through heating more difficult to identify, especially for populations born in a highly turbulent ISM.

The second mode of radial redistribution, which \cite{SellwoodBinney2002} introduced, is via scattering at the corotation resonance of transient spiral arms, which can drive permanent changes in a star's $j_\phi$ without increasing its random orbital energy.
This mechanism allows stars to change guiding-center radius while preserving orbital circularity.
Later work showed that resonant overlap between multiple patterns, such as spiral-spiral or bar-spiral interactions, can drive large changes in $j_\phi$ \citep{MinchevFamaey2010, Minchev2012b}, as well as the formation and subsequent slowing of the bar \citep{Chiba2021, Haywood2024, Khoperskov2024}. Together, these processes provide a mechanism for stars born on nearly circular orbits to move across the disk while largely preserving the shape of their orbits, making it difficult to detect in the orbits of stars today.

We now define key terminology that we use, given that previous works have used varying nomenclature.
We use `radial redistribution' for \textit{any} change in galactocentric radius or orbital angular momentum.
Some works refer to this as `radial migration', but other works use `radial migration' specifically to refer to corotation-resonance scattering (as in the previous paragraph), so we use `radial redistribution' as the more general term.
We refer to `heating' as any \textit{increase} in eccentricity (or radial action, $j_R$), which is also called `blurring' \citep{SchoenrichBinney2009a}.
By contrast, `cooling' is a \textit{decrease} in orbital eccentricity (or $j_R$).
We refer to `cold torquing' as: a net change in orbital angular momentum that preserves eccentricity/circularity (no significant change in $j_R$); this is also called `churning' \citep{SchoenrichBinney2009a}.
Our use of `cold torquing' is inspired by \citet{Daniel2019}, though they used the term specifically to refer to the kinematically cold changes in angular momentum that occur at the corotation-resonance (as in the previous paragraph), while we refer to it more generally as any change in $j_\phi$ that preserves eccentricity, regardless of the specific dynamical process.
Furthermore, throughout this work we measure \textit{net} changes in $j_\phi$ and eccentricity for a star particle, between its birth and today, regardless of its detailed dynamical history.
We are motivated by connections to galactic archeology, to infer the birth conditions of populations of stars based on their properties today.
Thus, we generally refer to a population of stars today as being `cold torqued', to emphasize that we measure the net change of a star particle from birth to today.

The extent to which cold torquing has shaped the stellar distributions we observe today remains poorly understood.
Observational works have attempted to infer the prevalence of cold torquing in the MW \citep[for example][]{Frankel2020, Feltzing2020, Lehmann2024, Lian2024}, but this generally requires assumptions about the past state of the disk, including the birth orbits of stars, the evolution of the ISM metallicity and its radial gradient, and the nature of the perturbations that drove redistribution. Such assumptions directly affect the inferred strength and consequences of cold torquing. This is because the present-day structure of a galaxy reflects both the conditions of the star-forming ISM at birth and the subsequent dynamical evolution of its stars, and these two effects are difficult to disentangle observationally. Different assumptions about the past disk can therefore lead to different conclusions about the importance of cold torquing. For example, some works have claimed that old stars have undergone substantial radial redistribution, while the MW has remained dynamically cool, placing strong constraints on mechanisms that redistribute angular momentum without excessive heating \citep{Frankel2020, Hamilton2024}. Similarly, \cite{SchoenrichBinney2009a} showed that the inferred strength of cold torquing can depend strongly on the assumed metallicity radial gradient of the ISM, with steeper gradients requiring less cold torquing to explain the same observations. Theoretical studies are therefore essential for understanding how efficiently cold torquing operates in disks.

Several theoretical works (using analytic models, idealized $N$-body simulations, idealized hydrodynamic simulations) have shown that stars can undergo substantial changes in $j_\phi$ through interactions with transient spiral structure \citep[for example][]{SellwoodBinney2002, Roskar2012, VeraCiro2014}, bar-spiral resonance overlap \citep[for example][] {Minchev2012b}, and slowing bars \citep[for example][]{Halle2018, Khoperskov2020, Chiba2021}, while preserving orbital circularity. Other theoretical works have claimed that cold torquing is most efficient for dynamically colder populations, which are able to couple most strongly to perturbations like corotation resonances of spiral arms \citep[for example][]{ VeraCiro2014, Daniel2018, Daniel2019}. Similarly, \cite{Solway2012} found that while cold torquing is most effective in thin disks, migration efficiency decreases only gradually with increased vertical motion, meaning thick disk stars can still couple to spiral perturbations and undergo radial redistribution. 
This is also consistent with \cite{Beraldo2020}, who found that thick disk stars can undergo cold torquing, with migration efficiency depending more strongly on orbital eccentricity than on vertical excursion.

Thus, most prior theoretical works on cold torquing have used idealized models/simulations, which make assumptions about the state of the MW at earlier times, typically assuming that all stars were born on (near) circular orbits, that they evolved in a relatively thin disk, and that they interacted mainly with internal structures like spirals, bars, and GMCs. Such assumptions are unlikely to hold throughout galaxy assembly, when disks were thicker, more turbulent, and less dynamically settled than today. Therefore, the thin well-ordered disks observed in the local universe are not generally representative of galaxies at earlier stages of their history.

For this reason, understanding the full nature of radial redistribution requires simulations that follow the cosmological evolution of the galaxy, including disk assembly and settling, along with the evolution of gas accretion, star formation, and mergers. Cosmological zoom-in simulations are especially well suited for this problem because they self-consistently capture both the evolving, clumpy, multiphase ISM and the internal and external perturbations that drive radial redistribution over cosmic time. Only a few works have examined radial redistribution, typically using a small number of cosmological simulations \citep{Blazquez2009, Martig2014, Grand2016a, Grand2016b, Vincenzo2020, verma2021, Lu2022, Boecker2022, Okalidis2022, Dubay2024, Minchev2025, wiggins2025, Bernaldez2026}. These studies have generally quantified the extent and consequences of redistribution, including its effects on galaxy structure, vertical disk heating, and elemental abundance evolution. Some have also discussed possible drivers such as bars and spiral structure. However, these works typically have not isolated the underlying dynamical channels of redistribution \citep[but see][]{wiggins2025}.

Few studies directly quantified the degree to which radial redistribution proceeded dynamically hot or cold. Among the works above, \cite{Grand2016a} examined \textit{vertical} disk heating associated with radial redistribution, including cases in which stellar populations become vertically cooler with time, but they did not examine changes to orbital eccentricity. More generally, vertical heating is related to, but distinct from, the question of whether radially-redistributed stars increased, maintained, or decreased their orbital eccentricities. A small number of studies used idealized, non-cosmological simulations to explore stellar orbits becoming more circular since birth \citep{Khoperskov2020, Struck2026}, but no works have examined this in a cosmological context.

\cite{Bellardini2026} quantified the total amount of stellar radial redistribution, from birth to today, across 11 MW-mass galaxies from the FIRE-2 cosmological simulations.
They examined changes in orbital radius, angular momentum, and azimuthal velocity. They showed that $\sigma(\Delta R_{\rm orbit})$, a common metric for the strength of radial redistribution, increases with age up to $\approx 3 \Gyr$ and then saturates at $\approx 2 \kpc$, and that this is largely consistent with recent inferences of the MW over this age range \citep{Frankel2020, Ratcliffe2025a}. However, \cite{Bellardini2026} did not quantify how `hot' or `cold' radial redistribution is, from birth to today, which is the focus of our analysis.
Our work is, to our knowledge, the first to quantify how dynamically `hot' or `cold' is the radial redistribution (via torquing) of stellar orbits in cosmological simulations of MW-mass galaxies that model the key processes and structures expected to drive radial redistribution and heating.
Specifically, we quantify how orbital eccentricity has changed from birth to today, for stars that underwent significant radial redistribution (torquing), in 12 MW-mass galaxies from the FIRE cosmological zoom-in simulations, by measuring the fraction of star particles that are cold torqued since birth.

\section{Methods}

\subsection{FIRE Simulations}

We use cosmological zoom-in simulations from the Feedback in Realistic Environments (FIRE) project\footnote{
FIRE project web site: \href{http://fire.northwestern.edu}{http://fire.northwestern.edu}}, which are publicly available \citep{Wetzel2023, Wetzel2025}.
All but one of our simulations use the FIRE-2 model \citep{Hopkins2018b}, and we introduce one new simulation, m12q, which uses the newer FIRE-3 model \citep{Hopkins2023a}.
Our sample contains $6$ cosmologically isolated MW/M31-mass galaxies from the \textit{Latte} suite \citep[introduced in][]{Wetzel2016} and $6$ Local Group-like MW$+$M31 pairs from the `ELVIS on FIRE' suite \citep{GarrisonKimmel2019a, GarrisonKimmel2019b}. These systems have halo masses of $\Mthm = 1 - 2 \times 10^{12} \Msun$, where $M_{\rm 200m}$ denotes the total mass within the radius whose mean density is $200$ times the mean density of the Universe. We omit $3$ galaxies from the \textit{Latte} suite: m12r and m12z because of their low stellar masses, and m12w because of its unusually compact disk.
Table~\ref{table:galaxies} lists the galaxies included in our analysis together with their key properties today.

These baryonic cosmological simulations follow the evolution of stars and gas, along with dark matter, in a high-resolution zoom-in region placed within a lower-resolution cosmological volume. The initial conditions are set within periodic cosmological boxes with side lengths of $70.4 - 172 \Mpc$ and were generated at $z \approx 99$ using \textsc{MUSIC} \citep{HahnAbel2011}. Each simulation outputs 600 snapshots to $z = 0$, with a typical spacing of $\approx 25 \Myr$. All simulations assume a flat $\Lambda$CDM cosmology with parameters broadly consistent with \citet{Planck2020b}: $h = 0.68 - 0.71$, $\Omega_{\Lambda} = 0.69 - 0.734$, $\Omega_{\rm m} = 0.266 - 0.31$, $\Omega_{\rm b} = 0.0455 - 0.048$, $\sigma_{\rm 8} = 0.801 - 0.82$, and $n_{\rm s} = 0.961 - 0.97$.

The \textit{Latte} suite has an initial baryonic particle mass of $7070 \Msun$.
However, the typical mass of a star particle is $\approx 5000 \Msun$ because of stellar mass loss.
A dark matter particle mass resolution of $3.5 \times 10^5 \Msun$. The ELVIS simulations have roughly twice better mass resolution. Thelma \& Louise has initial baryonic particle masses of $4000 \Msun$, while Romeo \& Juliet and Romulus \& Remus has initial baryonic particle masses of $3500 \Msun$. The gravitational force softening for star and dark matter particles is fixed at Plummer-equivalent values of $\epsilon_{\rm star} = 4 \pc$ and $\epsilon_{\rm dm} = 40 \pc$, with force softenings comoving at $z > 9$ and physical thereafter. For gas cells, the force softening is adaptive and follows the hydrodynamical kernel smoothing, reaching a minimum of $1 \pc$.
In the typical ISM, at densities of $\approx 1 \cci$, the softening/smoothing is $\approx 40 \pc$.

All simulations used the Meshless Finite Mass hydrodynamics method \citep{Hopkins2015}, and all simulations other than m12q used the FIRE-2 model for star formation and stellar feedback \citep{Hopkins2018b}. Gas cells undergo metallicity-dependent radiative heating and cooling across a temperature interval of $10-10^{10}$K, with contributions from free-free, photoionization and recombination, Compton, photoelectric and dust collisional, cosmic ray, molecular, metal-line, and fine-structure processes. The simulations follow $11$ elements (H, He, C, N, O, Ne, Mg, Si, S, Ca, Fe) and include a spatially uniform, time-dependent cosmic ultraviolet background from \citet{FaucherGiguere2009}.

Star formation occurs in gas that is dense, self-gravitating, Jeans-unstable, cold ($T < 10^{4}$K), and self-shielding and molecular \citep[following][]{KrumholzGnedin2011}. Once a gas cell satisfies these criteria, it probabilistically converts into a star particle on its local free-fall time.
Each star particle inherits the mass and metallicity of its progenitor gas cell and represents a single stellar population, assuming a \citet{Kroupa2001} initial mass function. The FIRE simulations include the major channels of stellar feedback, including continuous mass loss from stellar winds, core-collapse and white-dwarf supernovae, radiation pressure, photoionization, and photoelectric heating. In FIRE-2, stellar winds and their yields follow a combination of models \citep{vandenHoek1997, Marigo2001, Izzard2004} synthesized in \citet{Wiersma2009}, while core-collapse supernova rates come from \textsc{Starburst99} \citep{Leitherer1999}, and the white-dwarf supernova rates from \citet{Mannucci2006}. We adopt \citet{Nomoto2006} and \citet{Iwamoto1999} for the nucleosynthetic yields of core-collapse and white-dwarf supernovae, respectively.
All simulations explicitly model subgrid (unresolved) turbulent diffusion of metals in gas \citep{Hopkins2017, Su2017, Escala2018}.

We also include m12q, a resimulation of a galaxy originally presented in \citet{Hopkins2014a}, and also analyzed at lower resolution in \citet{barry2026}.
The new version we analyze here is at higher resolution and uses the FIRE-3 model \citep{Hopkins2023a}. FIRE-3 differs from FIRE-2 primarily through modest updates to various physical models, including the stellar evolution models, nucleosynthetic yields, low-temperature ISM cooling and chemistry, and the metagalactic ultraviolet background \citep{FaucherGiguere2020}.
Consistent with the FIRE-2 simulations we analyze, this FIRE-3 version of m12q does not include magnetohydrodynamics.
These updates are important for various predictions of stellar abundances, gas cooling, detailed ISM structure, and related observables, but as \cite{Hopkins2023a} showed, they have only modest effects on galaxy-wide properties, including galaxy-wide dynamics, and most conclusions from FIRE-2 are unlikely to change dramatically.
Thus, we do not expect the differences between the FIRE-2 and FIRE-3 models to qualitatively alter the galaxy-wide dynamical quantities we analyze.
We include m12q in our analysis primarily because, as Table~\ref{table:galaxies} shows, it is a particularly early-forming galaxy, and many works argue that the MW is particularly early-forming \citep[for example][]{BelokurovKravtsov2022, Conroy2022, Xiang2022}.

Crucial for our analysis, the FIRE simulations self-consistently model the multiphase ISM, including overdensities that can scatter stellar orbits and drive radial redistribution. For example, they resolve GMCs \citep{Benincasa2020, Guszejnov2020}, the formation of (massive) star clusters \citep{Ma2020a, Grudic2023, Sameie2023, Bhattarai2024}, spiral arms \citep{Orr2023, Quinn2025}, and bars \citep{Ansar2025}. The presence of these structures, combined with cosmological effects including gas accretion, galaxy mergers, and satellite interactions, makes it possible to investigate stellar cold torquing in a self-consistent cosmological context without adopting assumptions or analytic approximations about how these structures evolve or influence the galaxy.

In particular, \cite{McCluskey2025} provided important context for our work.
They compared the stellar velocity dispersions of these FIRE-2 MW-mass galaxies to the MW and nearby disk galaxies. They found that these FIRE-2 simulations broadly agree with M31, M33, and a sample of 19 PHANGS galaxies, but they are dynamically hotter than the MW by roughly a factor of two at most stellar ages.
Thus, FIRE-2 is reasonably representative of nearby disk galaxies, while the MW is unusually dynamically cold.

\subsection{Measuring Disk Dynamics}

We characterize stellar orbits using cylindrical coordinates. For each snapshot, we first determine the galaxy center using an iterative zoom-in method on all star particles that end up within the MW-mass galaxy today \citep[see][]{Wetzel2023}. We then measure positions and velocities relative to this center separately at each snapshot. To set the orientation of the coordinate system, we define the disk plane from the moment-of-inertia tensor of the youngest 25\% of stars that together contain 90\% of the stellar mass within 10 kpc at each snapshot. We use young stars to trace the disk orientation, as opposed to using all stars and gas, because older stellar populations are generally not disky, and the morphology of gas can be highly variable, especially during episodes of strong feedback.

In this coordinate system, we define the galactocentric radius $R$ as the cylindrical distance from the galaxy center and $Z$ as the height above or below the disk midplane. For each star particle, we measure the azimuthal velocity $v_\phi$ and the specific angular momentum $j_\phi = R v_\phi$. 

Throughout, we quote $v_\phi / \sigma_{v,\rm 3D}$ as a measure of rotational support of a disk. For a given stellar population, $v_\phi$ is the median azimuthal velocity across a given radial range. We define $\sigma_{v, \rm 3D} = \sqrt{\sigma_{v,R}^2 + \sigma_{v,\phi}^2 + \sigma_{v,Z}^2}$, where $\sigma_{v,R}$, $\sigma_{v,\phi}$, and $\sigma_{v,Z}$ are the standard deviations of the radial, azimuthal, and vertical velocity components, respectively. For the purpose of this paper, we are interested in the dynamical state of the overall disk, so we measure the velocities and dispersions disk-wide, rather than in local patches. This distinction is important, because disk-wide velocity dispersions include coherent large-scale motions from non-axisymmetric structures, warps, and other perturbations, and therefore are significantly larger than velocity dispersions measured in local patches \citep[see][]{McCluskey2025}.

\subsection{Spatial Selection of Stars}

In analyzing each galaxy, we generally first select stars that formed \textit{in situ}, following \cite{Bellardini2022} ($d_{\rm birth} < 30 \kpc$ comoving), and located today at $|Z_{\rm now}| < 3 \kpc$ and $R_{\rm now} = 2 - 16 \kpc$, except in figures that show trends versus radius at birth or radius today.
Our motivation is to focus on stars in the disk region and avoid the more complex dynamics in the inner galaxy (bulge region).
This spatial selection is broadly consistent with previous related analyses of FIRE simulations \citep{McCluskey2024, Bellardini2026}.
We verified that our results are insensitive to this choice by reproducing key figures selecting stars instead using only the \textit{in situ} cut, as well as stars that formed within $R_{\rm birth} = 2 - 16 \kpc$, and we find negligible differences in the resulting trends between the three radial selections. 

Unless otherwise stated, we do \textit{not} impose any selection on the dynamical state (velocities, eccentricities, etc.) of the star particles that we analyze.

\subsection{Eras of Disk Formation}

\cite{McCluskey2024} described disks in FIRE as forming and evolving through 3 distinct eras: (1) a \textit{pre-disk} era, when stars formed on largely dispersion-dominated orbits; (2) an \textit{early-disk} era, characterized by the emergence of a dynamically hot, rotation-dominated thick disk ($v_\phi / \sigma_{v,\rm 3D} > 1$), during which disk settling/thinning occurs; and (3) a \textit{late-disk} era, marked by the formation of stars on thin-disk, near-circular orbits ($v_\phi / \sigma_{v, \rm 3D} \gtrsim 3$).
To be consistent with our spatial selection, we recompute these disk transition times using stars born within $R_{\rm birth} = 2 - 12 \kpc$.
Thus, our disk transition times differ slightly from \cite{McCluskey2024} because of the slight difference in radial selection.

While the three-era framework follows \cite{McCluskey2024}, we introduce an additional specific kinematic threshold to define the transition to the late-disk era. We choose the lookback time corresponding to the transition between the early- and late-disk phases to be when $v_\phi / \sigma_{v, \rm 3D} \gtrsim 3$. When examining the median $v_\phi / \sigma_{v, \rm 3D}$ with stellar age, for most galaxies the ratio increases monotonically toward younger ages until $v_\phi / \sigma_{v, \rm 3D} \gtrsim 3$, at which the trend flattens. The exception to this is galaxies that have not yet transitioned to the late-disk era. This flattening indicates that the thin disk has become fully settled, such that rotational support is saturated. 

The transition from early- to late-disk coincides closely with the transition from bursty to smooth star formation identified in \cite{Yu2021}. As Table~\ref{table:galaxies} shows, the typical difference in age between $t_{\rm lb}(v_\phi / \sigma_{v, \rm 3D} > 3)$ and $t^{\rm burst}_{\rm lb}$ is $\approx 0.23 \Gyr$, the maximum difference is $\approx 0.49 \Gyr$ and the minimum $\approx 0.02 \Gyr$. Despite being defined using an independent kinematic diagnostic, this suggests a physical connection between disk settling and the regulation of star formation.

We divide stars into 3 disk eras to examine how their orbital properties evolve under different dynamical birth conditions. Pre-disk stars formed before the disk formed, when the interstellar medium was highly turbulent, and rotation was not yet ordered. They therefore formed on eccentric, dispersion-dominated orbits and remained dynamically hot, making them the population most susceptible to large changes in angular momentum. Early-disk stars formed during an era when the disk already had ordered rotation, but velocity dispersions were still high; they exhibit the broadest distribution of birth eccentricities and also show substantial radial redistribution. Late-disk stars formed after the disk developed ordered rotation, and most were born on near-circular orbits. In general, we examine trends as a function of age, but in many figures, we show trends with age by breaking into these three eras and show that each era exhibits distinct behavior.

\begin{table*}
\centering
\caption{properties today of the 12 FIRE-2 galaxies we analyze}
\label{table:galaxies}
\begin{tabular}{|l|ccccc|}
\hline 

\textbf{galaxy} & 
$M^\star_{\rm 90}$&
\textbf{$t_{\rm lb} \left[\frac{v_\phi}{\sigma_{v,\rm 3D}} > 1 \right]$} &
\textbf{$t_{\rm lb} \left[\frac{v_\phi}{\sigma_{v,\rm 3D}} > 3\right]$} &
\textbf{$t^{\rm burst}_{\rm lb}$} &
$\frac{v_\phi}{\sigma_{v, \rm 3D}}$ now   \\ [2mm]

\hline

m12q   & 5.1  & 11.00 & 5.75 & - & 7.5    \\[1mm]
Romeo  & 5.9  & 10.25 & 6.50 & 6.52 & 6.3  \\[1mm]
m12m   & 10.0 & 9.25  & 3.75 & 3.81 & 5.8 \\[1mm]
m12b   & 7.3  & 7.75  & 6.75 & 6.32 & 4.9   \\[1mm]
Remus  & 4.0  & 7.75  & 5.75 & 5.88 & 5.6  \\[1mm]
Romulus & 8.0 & 7.50  & 5.25 & 4.90 & 4.6 \\[1mm]
Juliet & 3.3  & 7.50  & 4.00 & 4.40 & 4.6  \\[1mm]
m12f   & 6.9  & 7.25  & 5.50 & 5.01 & 3.8  \\[1mm]
Louise & 2.3  & 6.75  & 5.50 & 5.56 & 4.6  \\[1mm]
m12c   & 5.1  & 6.50  & 3.50 & 3.70 & 3.8  \\[1mm]
m12i   & 5.3  & 5.75  & 3.50 & 3.14 & 4.2   \\[1mm]
Thelma & 6.3  & 4.25  & 2.50 & 2.57 & 3.3   \\
\hline
\end{tabular}
\tablecomments{
Galaxies are in decreasing order of disk onset time.
Columns show: galaxy name; $M^\star_{90}$ is the stellar mass within $R^\star_{90}$; the lookback time corresponding to the onset of disk formation, when $v_{\rm \phi} / \sigma_{v, \rm 3D} > 1$ permanently for stars at birth; the lookback time corresponding to the onset of a rotationally dominated disk (late-disk onset time), defined as when $v_{\rm \phi} / \sigma_{v, \rm 3D} > 3$; the lookback time corresponding to the transition from bursty to smooth star formation \citep{Yu2021}; the present-day $v_{\phi} / \sigma_{v, \rm 3D}$, for stars younger than $100 \Myr$.
}
\end{table*}

\subsection{Measuring Orbital Eccentricity}
\label{measure_e}

We quantify how `hot' or `cold' each star particle's orbit is via orbital eccentricity, $e$, which characterizes the degree of non-circular motion.
We first compute the orbital circularity, $\eta$, of each star particle,
\begin{equation}
\eta = \frac{j_\phi}{j_{\mathrm{circ}}(E)} ,
\end{equation}
where $j_\phi$ is the orbital specific angular momentum (azimuthal action), and $j_{\mathrm{circ}}(E)$ is the specific angular momentum of a circular orbit with the same total energy, $E$.
We then define eccentricity from circularity\footnote{
A common definition of eccentricity is $e = (r_{\rm apo} - r_{\rm peri}) / (r_{\rm apo} + r_{\rm peri})$.
However, in realistic galactic potentials, eccentricities based on $r_{\rm apo}$ and $r_{\rm peri}$ are not necessarily well defined, because orbits are not necessarily closed, and a star does not necessarily reach the same radial extrema each cycle \citep[see][]{Eggen1962, JoBovy}.
Furthermore, realistic galaxy potentials as in our simulations are non-axisymmetric and evolve.
We therefore choose to measure eccentricity as $e = \sqrt{1 - \eta^2}$ \citep[see also][for example]{Wetzel2011, Vasiliev2022}, which is an exact relation in a Keplerian potential.
$e$ and $\eta$ both range from 0 to 1, but using eccentricity has the advantage of putting more of the dynamic range closer to a circular orbit, our regime of interest.
For reference, $e = 0.1, 0.2, 0.4, 0.6, 0.8$ correspond to $\eta = 0.995, 0.98, 0.92, 0.8, 0.6$, respectively.
Given this straightforward transformation, our results would be the same if we used thresholds in $\eta$ instead.
} as
\begin{equation}
e = \sqrt{1 - \eta^{2}}.
\end{equation}

To compute the total energy of a star particle, one must evaluate the gravitational potential at the particle's position.
Following \cite{Bellardini2026}, we compute $j_{\rm circ}(E)$ by first determining $R_{\mathrm{circ}}(E)$, the radius of a circular orbit that has the same total energy as the star particle, using:
\begin{equation}
\label{eqn:RE}
\Phi(R_{\rm circ}(E)) + \frac{1}{2} R_{\rm circ}(E) 
\left. \frac{{\rm d}\Phi(R)}{{\rm d}R} \right \rvert_{R = R_{\rm circ}(E)} = E.
\end{equation}
Here, $\Phi$ is the gravitational potential and $E$ is the total energy.
The FIRE simulations store the gravitational potential at the location of each particle at every snapshot. 
To estimate how the potential varies with galactocentric radius, $R$, we calculate a potential profile by averaging the potentials of all star, gas, and dark matter particles within $\pm 0.3 \kpc$ of the galactic midplane, in radial bins of $250 \kpc$ at every snapshot. Because Equation~\ref{eqn:RE} assumes a single value of $R$ for every $\Phi$, we require the gravitational potential to increase monotonically with radius. 
In rare cases, the potential decreases with $R$, given non-axisymmetric and time-dependent structures in the galaxy, particularly at early times.
Consistent with \cite{Bellardini2026}, to enforce monotonicity, we replace any non-monotonic value with that of the adjacent inner radial bin.
Using Equation~\ref{eqn:RE}, we then compute the total energy of a circular orbit as a function of $R$. We determine $R_{\rm circ}(E)$ by numerically solving for the radius at which the circular orbital energy equals the total energy of the star particle. Finally, we calculate $j_{\rm circ}(E)$ by first computing the circular velocity,
\begin{equation}
\label{eqn:vcirc}
v_{\rm circ} = \sqrt{R_{\rm circ}(E) \, \frac{{\rm d} \Phi}{{\rm d}R} \biggr \rvert_{R = R_{\rm circ}(E)}}
\end{equation}
and then evaluating 
\begin{equation}
\label{eqn:jcirc}
j_{\mathrm{circ}}(E) = v_{\mathrm{circ}} \, R_{\mathrm{circ}}(E).
\end{equation}

The Romulus \& Remus simulation did not store the gravitational potential at the particle locations. We therefore estimate the potential using a spherical enclosed-mass approximation. At each snapshot, we calculate the enclosed mass profile, $M(< R)$, and approximate the radial derivative of the potential as
\begin{equation}
\frac{{\rm d}\Phi}{{\rm d}R} = \frac{G M_{\rm tot}(< R)}{R^2}.
\end{equation}
We then reconstruct $\Phi(R)$ by numerically integrating ${\rm d}\Phi / {\rm d}R$ inward from the outermost radial bin. Using this reconstructed potential, we follow the same procedure described above to determine $R_{\rm circ}(E)$ from Equation~\ref{eqn:RE}. We then evalulate the circular velocity directly from the enclosed mass profile using
\begin{equation}
v_{\rm circ} =
\sqrt{\frac{G M_{\rm tot}(< R)}{R}}.
\end{equation}
We use this to construct $j_{\rm circ}(R)$, and then evaluate this at $R_{\rm circ}(E)$ to obtain $j_{\rm circ}(E)$.

This method introduces an additional caveat, because the enclosed-mass method assumes a spherical potential and therefore does not capture the full non-axisymmetric or flattened structure of the galaxy. To assess the size of this effect, we tested the stellar eccentricities calculated via the two methods for Romeo. Comparing the enclosed-mass eccentricities to those measured using the recorded potential, we find a mean offset of $-0.023$ for present-day eccentricity and $0.0016$ for birth eccentricity. The typical mean offset is therefore small, but the enclosed-mass approximation can introduce differences on a star-particle-by-star-particle basis. Importantly, this does not add substantial scatter to the distributions of birth eccentricity, present-day eccentricity, or $\Delta e$. We also repeated our analysis both including and excluding Romulus \& Remus: their inclusion does not shift any of our general trends. We therefore include them, because they increase our sampling of varied formation histories, particularly because they formed in a Local Group-like environment.

\subsection{Measuring Radial Redistribution}

We measure the amount of radial redistribution, and the change in orbital eccentricity, for each star particle from its birth to today.
We quantify radial redistribution using fractional changes in angular momentum, $\Delta j_\phi / |j_{\phi,\mathrm{birth}}|$, rather than changes in orbital radius.
This is because our key goal is to quantify the degree to which redistribution via torquing (sometimes called `churning') is typically dynamically `hot' or `cold', between birth to today.
Thus, we do not consider radial redistribution caused by radial heating (sometimes called `blurring'), which increases eccentricity without changing angular momentum. As a result, our measurements represent a lower limit on the total radial redistribution experienced by each population \citep[see][]{Bellardini2026}. Negative values indicate inward migration. However, stars with $\Delta j_\phi / |j_{\phi,\rm birth}| < -1$ have crossed through retrograde orbits and ultimately migrate outward. We therefore classify them as outward migrators along with those that have $\Delta j_\phi / |j_{\phi,\rm birth}| > 0$. This distinction is important because, on average, 32\% of star particles born in the pre-disk era, and 2\% of star particles born in the early-disk era, formed retrograde.

For all of the analysis that follows, we tested our results using different radial redistribution metrics: the instantaneous physical radius, $R$, and the radius of a circular orbit with the same angular momentum $R_{\rm circ}(j)$ (sometimes referred to as the `guiding-center radius'):
\begin{equation}
\label{eqn:Rj}
R_{\rm circ}(j) = j^{2/3} \left[ \frac{\,d\Phi(R)}{\,dR} \biggr\rvert_{R = R_{\rm circ}(j)} \right] ^{-1/3} .
\end{equation}
\cite{Bellardini2026} provided a comprehensive comparison of 5 definitions of orbital radius and showed that consistent trends in redistribution emerge when one measures radius today and changes in radius self-consistently. Appendix~\ref{sec:rr_metrics} shows the median change in eccentricity from birth to today, $\Delta e$, for populations of stars born in each of the three eras as a function of 6 radial redistribution metrics. The two metrics that use galactocentric radius, $R$, produce systematically larger and almost exclusively positive changes with $\Delta e$, demonstrating how sensitive changes to $R$ are to eccentricity.

\begin{figure}[ht!]
\includegraphics[width = 0.975 \linewidth]{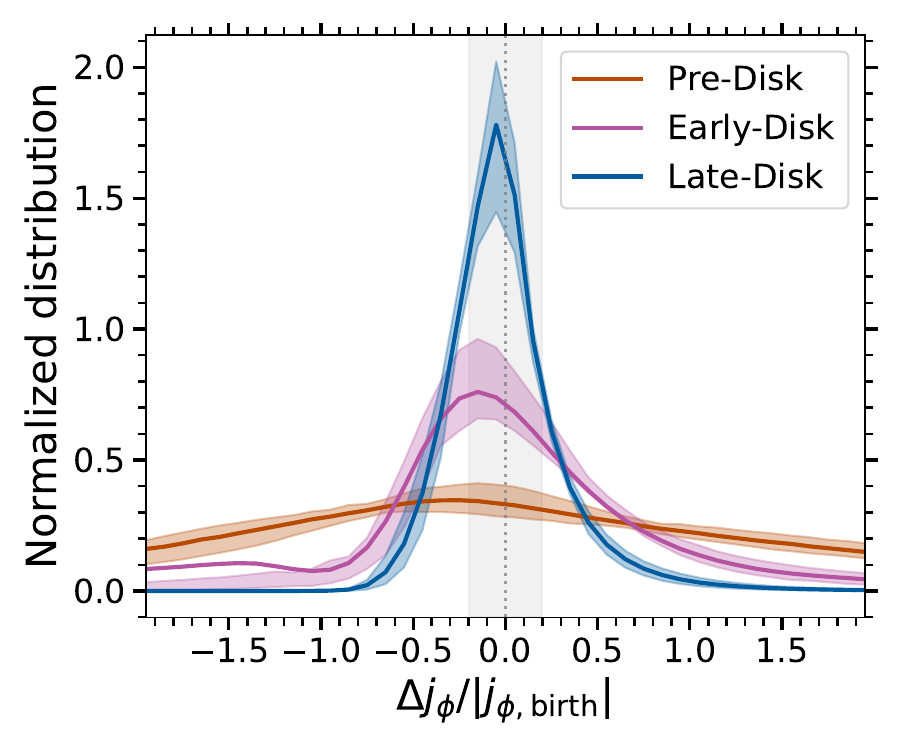}
\caption{
Normalized distribution of the fractional change in orbital angular momentum, $\Delta j_\phi / |j_{\rm \phi, birth}|$, for all stars (regardless of birth eccentricity).
We separate populations of stars into the three eras of disk evolution in which they were born.
We measure these eras separately for each galaxy, but they correspond to typical ages of: pre-disk ($\gtrsim 8 \Gyr$ ago), early-disk ($\approx 4 - 8 \Gyr$ ago), and late-disk ($\lesssim 4 \Gyr$ ago).
Lines show the mean, shaded regions show the 68th percentile scatter across the 12 galaxies.
The amount of radial redistribution in terms of the width of the distribution, $\sigma(\Delta j_\phi / |j_{\rm \phi, birth}|)$, generally increases with age.
Throughout, we define `significant' radial redistribution as $|\Delta j_\phi / j_{\rm \phi,birth}| > 0.2$, which includes stars beyond the grey shaded region.
}
\label{j_dist}
\end{figure}

Figure~\ref{j_dist} shows the normalized distribution $\Delta j_\phi / |j_{\phi,\rm birth}|$, for star particles that formed in the three eras of disk evolution. Consistent with \cite{Bellardini2026}, all eras exhibit a slight preference for inward migration.
The width of the distribution, $\sigma(\Delta j_\phi / |j_{\phi, \rm birth}|)$, generally increases with age. Pre-disk stars display the broadest distribution. Early- and late-disk star particles are more narrowly distributed and are roughly symmetric around zero. These trends show that radial redistribution occurs in all eras, is most prominent for the dynamically hot pre-disk population, and is mildly biased toward inward migration. We define that a star particle has radially redistributed `significantly' if $\left| \Delta j_\phi / j_{\phi,\mathrm{birth}} \right| > 0.2$. Figure~\ref{j_dist} motivates this choice: 7\% of pre-disk star particles, 26\% of early-disk star particles, and 57\% of late-disk star particles all lie within $\left| \Delta j_\phi / j_{\phi,\rm birth} \right| < 0.2$, indicating that a substantial fraction of star particles, particularly at late times, experience only modest changes in angular momentum. By adopting this threshold, we explicitly select star particles that have undergone significant changes in angular momentum and exclude the population whose radial redistribution is comparatively weak.

In summary, for each star particle, between its birth and today:
\begin{itemize}
    \item We quantify radial redistribution as
    \[
    \Delta j_\phi / |j_{\phi,\mathrm{birth}}| \equiv (j_{\phi,\mathrm{now}} - j_{\phi,\mathrm{birth}}) / |j_{\phi,\mathrm{birth}}|.
    \]

    \item \emph{Significant} radial redistribution is
    \[
    \left| \Delta j_\phi / j_{\phi,\mathrm{birth}} \right| > 0.2.
    \]

    \item \emph{Inward migrators} satisfy
    \[
    -1 < \Delta j_\phi / |j_{\phi,\mathrm{birth}}| < 0.
    \]

    \item \emph{Outward migrators} satisfy
    \[
    \Delta j_\phi / |j_{\phi,\mathrm{birth}}| > 0
    \quad \text{or} \quad
    \Delta j_\phi / |j_{\phi,\mathrm{birth}}| < -1.
    \]

\end{itemize}

\section{Results}

\subsection{Changes to Eccentricity}

\begin{figure}[ht!]
\includegraphics[width = 0.975 \linewidth]{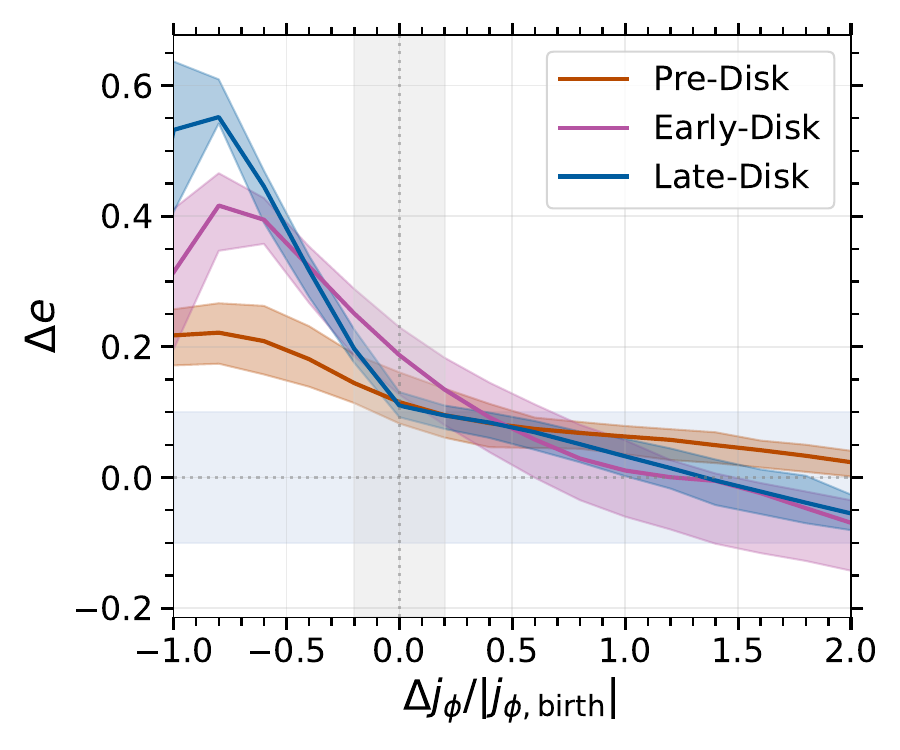}
\caption{
Median change in eccentricity, from birth to today, versus amount of radial redistribution (via fractional change in orbital angular momentum), for all stars (regardless of birth eccentricity).
We separate populations of stars into the three eras of disk evolution in which they formed.
We measure these eras separately for each galaxy, but they correspond to typical ages of: pre-disk ($\gtrsim 8 \Gyr$ ago), early-disk ($\approx 4 - 8 \Gyr$ ago), and late-disk ($\lesssim 4 \Gyr$ ago).
Lines show the mean, shaded regions show the 68th percentile scatter across the 12 galaxies.
The blue horizontal shaded region shows stellar populations that we define as staying dynamically `cold': $|\Delta e| < 0.1$.
The amount of dynamical heating decreases monotonically with $\Delta j_\phi / |j_{\rm \phi, birth}|$, such that outward-migrating stars are much more likely to remain cold.
In particular, stars born in the late-disk era (blue) show a dramatic rise in $\Delta e$ below $\Delta j_\phi / |j_{\rm \phi, birth}| = 0$.
The vertical grey shaded region shows stars that have not significantly redistributed, and therefore we generally do not include them in our analysis.
That said, even that non-redistributed population of stars tends to be dynamically heated, with an average $\Delta e \approx 0.1 - 0.2$.
The only stars with $\Delta e \approx 0$ on average are those that redistributed outward significantly, with $\Delta j_\phi / |j_{\rm \phi, birth}| \approx 1$.
}
\label{dele_j}
\end{figure}

Understanding how stellar orbital eccentricities evolve is central to interpreting both radial redistribution and the origin of dynamically cold stellar populations. We first examine the changes in eccentricity from birth to today and how they vary with the degree of radial redistribution experienced. We then use the distributions of eccentricity at birth and today to provide additional context for these changes, thereby establishing a baseline picture of how stellar orbits evolved across each dynamical phase of the disk and motivating the subsequent analysis of changes in eccentricity.

Figure~\ref{dele_j} shows the median change in eccentricity, $\Delta e$, versus the fractional change in orbital angular momentum, $\Delta j_\phi / |j_{\phi,\rm birth}|$, for stars that formed in the three eras of disk evolution.
Across all disk populations, $\Delta e$ increases with progressively greater inward migration ($\Delta j_\phi / |j_{\phi,\mathrm{birth}}| < 0$), such that more negative values correspond to larger increases in eccentricity. In contrast, $\Delta e$ decreases with increasing outward migration ($\Delta j_\phi / |j_{\phi,\mathrm{birth}}| > 0$). Although we do not show it, $\Delta e$ declines again for $\Delta j_\phi / |j_{\phi,\mathrm{birth}}| < -1$, consistent with the `normal' outward migrators. Overall, the amount of dynamical heating decreases monotonically with $\Delta j_\phi / |j_{\phi,\mathrm{birth}}|$, such that stars that redistributed outward are substantially more likely to remain dynamically cold. This behavior is particularly pronounced for stars born in the late-disk era, which exhibit a sharp rise in $\Delta e$ for $\Delta j_\phi / |j_{\phi,\mathrm{birth}}| < 0$. 

Contrary to the common assumption that radial redistribution is typically associated with heating, and that stars that remain near their birth places tend to preserve their initial orbits, Figure~\ref{dele_j} shows the opposite trend. Non-redistributed stars, lying within the vertical grey shaded region, on average experienced moderate heating, with $\Delta e \approx 0.1$–$0.2$. In contrast, the only population with $\Delta e \approx 0$ on average are stars that underwent significant outward redistribution, with $\Delta j_\phi / |j_{\phi,\mathrm{birth}}| \approx 1$.

\begin{figure*}[!ht]
\centering
\includegraphics[width = 0.975 \linewidth]{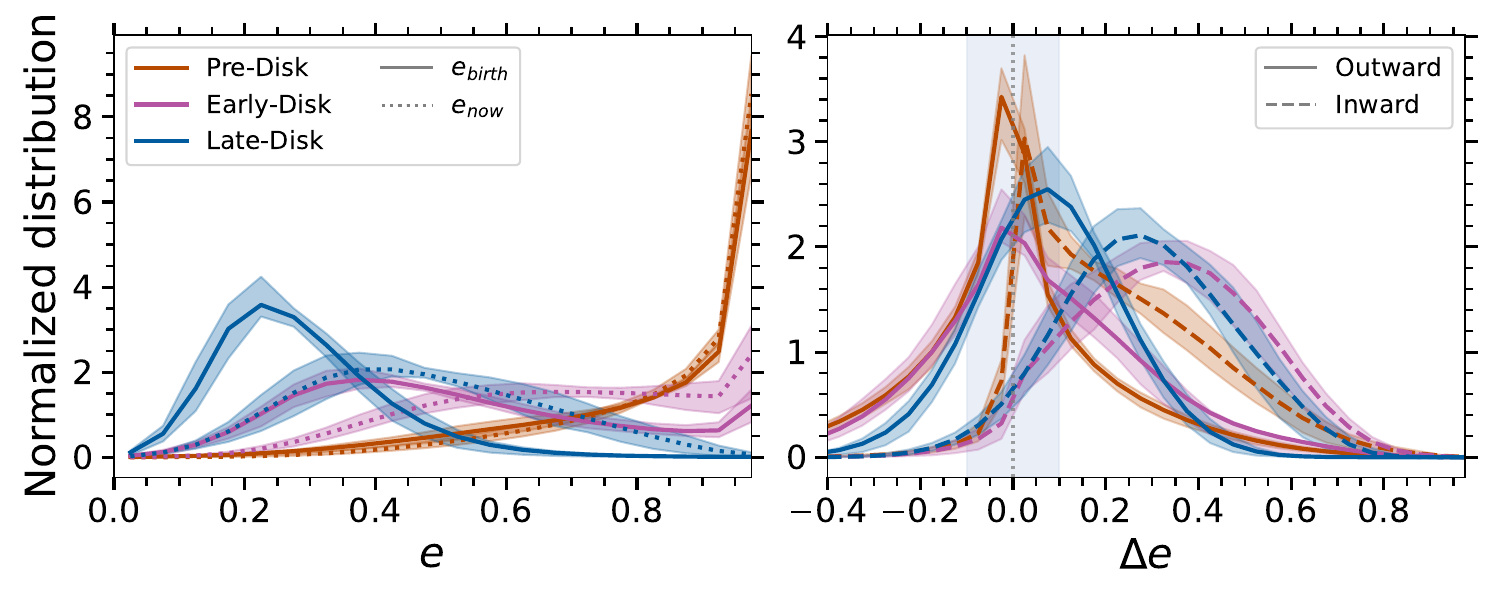}
\caption{
\textbf{Left}: Normalized distribution of eccentricity, $e$, at birth (solid) and today (dotted), for stars that experienced significant radial redistribution ($|\Delta j_\phi / j_{\rm \phi, birth}| > 0.2$).
We separate stars born in the three eras of disk evolution, which we measure separately for each galaxy: pre-disk (typically $e_{\rm birth} \gtrsim 0.9$), early-disk (typically $e_{\rm birth} \approx 0.4$), and late-disk (typically $e_{\rm birth} \approx 0.2$).
Lines show the mean, shaded regions show the 68th percentile scatter across the 12 galaxies.
\textbf{Right}: Normalized distribution of the change in eccentricity, $\Delta e = e_{\rm today} - e_{\rm birth}$, for all stars (regardless of birth eccentricity) that experienced significant radial redistribution, split by migration direction: outward-moving (solid) and inward-moving (dashed).
The blue vertical shaded region indicates stars that we consider remain dynamically `cold': $|\Delta e| < 0.1$.
Inward migrators are much more likely to be dynamically heated ($\Delta e \gtrsim 0.2$) than outward migrators, regardless of the dynamical era (age) of birth.
}
\label{dele_efen_dists}
\end{figure*}

While Figure~\ref{dele_j} quantifies the changes in orbital eccentricity, it does not show trends based on the orbits of stars at birth, or what eccentricities stars end up on today.
Thus, Figure~\ref{dele_efen_dists} (left) shows the distribution of eccentricity at birth (solid lines) and today (dotted lines) for stars born in each disk era that significantly radially redistributed.
Pre-disk stars formed on the most eccentric orbits ($e_{\rm birth} \approx 1$), reflecting the highly turbulent conditions of the early ISM. Early-disk stars exhibit the broadest range of birth eccentricities, consistent with a disk in transition toward ordered rotation but still characterized by high velocity dispersion, with an average $e_{\rm birth} \approx 0.4$. Late-disk stars formed on relatively circular orbits, with a peak near $e_{\rm birth} \approx 0.2$, showing that even the youngest stars do not form on perfectly circular orbits. These trends broadly align with previous analyses of the FIRE-2 simulations, which have shown that stellar kinematics at birth closely reflect the dynamical state of the ISM \citep[for example][]{Yu2021, Yu2023, Gurvich2023, McCluskey2024}. The left panel shows that our era-based classification captures the expected progression from a turbulent, dispersion-dominated disk to an increasingly rotationally supported disk.

Today, the eccentricity distributions of early- and late-disk stars systematically shift and broaden toward higher eccentricities, indicating substantial post-birth evolution. Late-disk stars today peak at $e_{\rm now} \approx 0.4$, while early-disk stars peak at $e_{\rm now} \approx 0.6$, demonstrating that, on average, stars that formed on relatively cold orbits became moderately heated over time. In contrast, the $e_{\rm now}$ distribution of pre-disk stars lies directly on top of their $e_{\rm birth}$ distribution, signifying little to no change in their eccentricity distribution since birth. This is because, independent of the specific mechanism, early- and late-disk stars have more available phase space to increase their eccentricities.

Figure~\ref{dele_efen_dists} (right) quantifies the change between $e_{\rm birth}$ and $e_{\rm now}$ for stars that migrated significantly outward (solid lines) or inward (dashed lines). This reinforces one of our central results from Figure~\ref{dele_j}: the direction of radial redistribution strongly influences how stellar eccentricities evolve, largely independent of the dynamical era the stars formed in. For pre-disk stars, outward migrators have $\Delta e$ symmetric around zero, while inward migrators are skewed toward small positive values. Early-disk stars exhibit the broadest distributions, with outward migrators having an average $\Delta e \approx 0$ and inward migrators having $\Delta e \approx 0.4$. Similarly, late-disk stars show outward migrators with an average $\Delta e \approx 0$ and inward migrators having typical values of $\Delta e \approx 0.2$. Across all three disk populations, inward migrators are more likely to dynamically heat, while outward migrators tend to remain cold. This inward–outward asymmetry motivates much of the analysis that follows: it suggests that we cannot understand changes in eccentricity solely in terms of age or dynamics at birth, but instead, we must interpret them in the context of the direction and (to a lesser degree) magnitude of radial redistribution.

\begin{figure*}[!ht]
\centering
\includegraphics[width = 0.975 \linewidth]{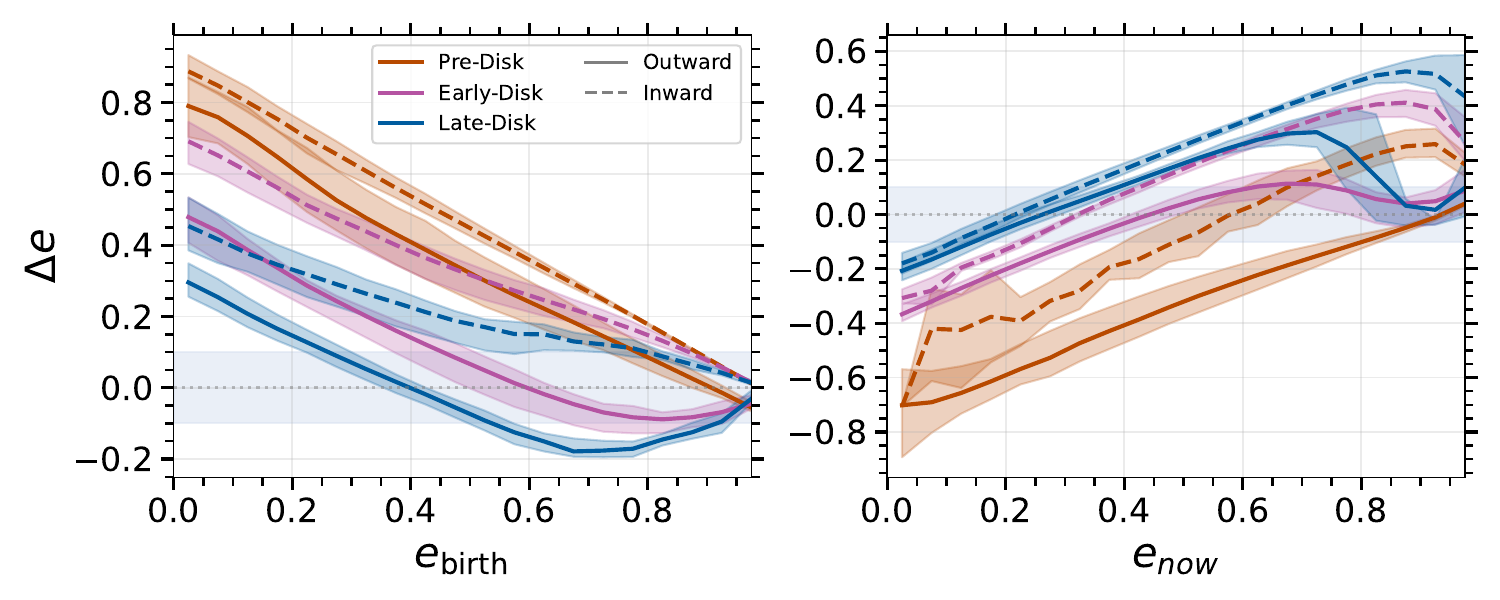}
\caption{
Median change in orbital eccentricity from birth to today, $\Delta e$, versus eccentricity at birth (left) and today (right), for all stars (regardless of birth eccentricity) that redistributed significantly inward (dashed) or outward (solid).
Different colored lines show stars born in the three eras of disk evolution.
Lines show the mean, shaded regions show the 68th percentile scatter across the 12 galaxies.
The blue horizontal shaded regions mark stars whose dynamical evolution from birth to today was `cold', with $|\Delta e| < 0.1$.
Stars born on near-circular orbits are the most susceptible to dynamical heating during radial redistribution.
Stars born with $e_{\rm birth} = 0.4 - 0.6$ (depending on era) are most likely to retain their birth eccentricity, and thus be cold torqued.
By contrast, stars today with $e_{\rm now} = 0.2 - 0.4$ (right) are most likely to have be cold torqued.
Stars on near-circular orbits today (right) arrived at near-circular orbits largely by dynamically cooling since birth, with $\Delta e \approx -0.2$ to $-0.7$, depending on the era.
}
\label{dele_efe}
\end{figure*}

Figure~\ref{dele_efe} shows the median change in eccentricity, $\Delta e$, versus eccentricity at birth (left panel) and today (right panel) for stars that redistributed significantly inward or outward. The left panel shows that, across all disk populations, inward migrators have systematically higher $\Delta e$ than outward migrators at fixed birth eccentricity. However, stars born on the most eccentric orbits converge toward $\Delta e \approx 0$, independent of disk era and migration direction. Almost no early- or late-disk stars were born at such extreme eccentricities. Only 7\% of early-disk stars and 0.07\% of late disk stars were born with $e_{\mathrm{birth}} > 0.9$. This behavior indicates that stars born dynamically hot largely preserve their initial eccentricities and remain hot, as they already occupy the upper range of available eccentricities and therefore cannot be heated further.

In contrast, early- and late-disk stars born on eccentric orbits ($\Delta e \approx 0.6-0.8$) on average dynamically cooled, with late-disk stars reaching down to $\Delta e \approx -0.2$. This suggests that stars born eccentric in a relatively colder disk more readily can cool when migrating outward. Stars born on low-eccentricity orbits tend to gain eccentricity, with pre-disk stars exhibiting the largest positive $\Delta e$, followed by early- and then late-disk stars. Essentially no pre-disk stars were born on circular orbits, consistent with the turbulent conditions of the early ISM. Taken together, these trends imply that stars born on near-circular orbits are the most susceptible to heating, especially for stars born in a turbulent and dispersion-dominated ISM. Additionally, stars born on moderately eccentric orbits ($e_{\rm birth} = 0.4 - 0.6$ depending on the era) are most likely to retain their birth eccentricity and thus be cold torqued.

Figure~\ref{dele_efe} (right) shows $\Delta e$ versus eccentricity today. Early- and late-disk stars on eccentric orbits today generally increased their eccentricities since birth. Outward-migrating pre-disk stars on the most eccentric orbits today exhibit $\Delta e \approx 0$. More generally, across $e_{\rm now} < 1$, outward-migrating pre-disk stars on average dynamically cooled, reflecting the fact that they were born on highly eccentric orbits. In contrast, stars on near-circular orbits today decreased their eccentricities since birth, with pre-disk stars reaching $\Delta e \approx -0.7$, early-disk stars $\Delta e \approx -0.4$, and late-disk stars $\Delta e \approx -0.2$. Pre-disk stars on near-circular orbits today are rare, $\approx 0.07\%$, but they occur in almost all FIRE-2 simulations \citep{Santistevan2021}. Throughout, inward migrators maintained systematically higher $\Delta e$ than outward migrators at fixed eccentricity today.

Figure~\ref{dele_efe} shows that the birth eccentricity corresponding to $\Delta e \approx 0$ on average does not match the present-day eccentricity associated with $\Delta e \approx 0$ on average. For example, in the late-disk population, $\Delta e \approx 0$ occurs at $e_{\mathrm{birth}} \approx 0.4$, whereas it corresponds to $e_{\mathrm{now}} \approx 0.2$. This difference reflects the underlying broad (and generally non-Gaussian) distributions of both quantities. The present-day eccentricity distribution is not simply a shifted version of the birth distribution, but instead results from the combined effects of the birth eccentricity distribution and the distribution of $\Delta e$. Because both distributions are broad and evolve across disk eras, the eccentricity at which $\Delta e \approx 0$ is most common does not correspond to identical peaks in $e_{\mathrm{birth}}$ and $e_{\mathrm{now}}$. In particular, relatively few stars are at extremely low eccentricities today, even if such stars are more likely to have experienced small $\Delta e$. As a result, the stars most likely to preserve their eccentricities depend on the full distribution of birth eccentricities and subsequent changes, rather than on a one-to-one correspondence between $e_{\rm birth}$ and $e_{\rm now}$.

Figure~\ref{dele_efe} shows that a star's birth eccentricity strongly influences whether it is more likely to heat up, cool down, or remain unchanged throughout radial redistribution. Stars born on near-circular orbits are the most likely to heat up, while stars born on very eccentric orbits tend to experience little net change in eccentricity. In contrast, outward-migrating early- and late-disk stars born on eccentric orbits are the most likely to cool down over time. Cooling, remaining cold, or experiencing only minimal heating occurs more frequently among stars that migrated outward, and the dynamical era in which a star forms plays a major role in determining how efficiently it avoids heating while undergoing radial redistribution. When considering present-day eccentricities, stars on near-circular orbits today become more circular, on average, since birth, whereas outward-migrating stars on the most eccentric orbits today were most likely born with similarly high eccentricities and inward migrators with high eccentricities likely got hotter since birth.

To summarize, the direction of radial redistribution and the dynamical era during which a star was born significantly and systematically affect how its orbit heats or cools over time.

\begin{figure}[ht!]
\includegraphics[width = 0.975 \linewidth]{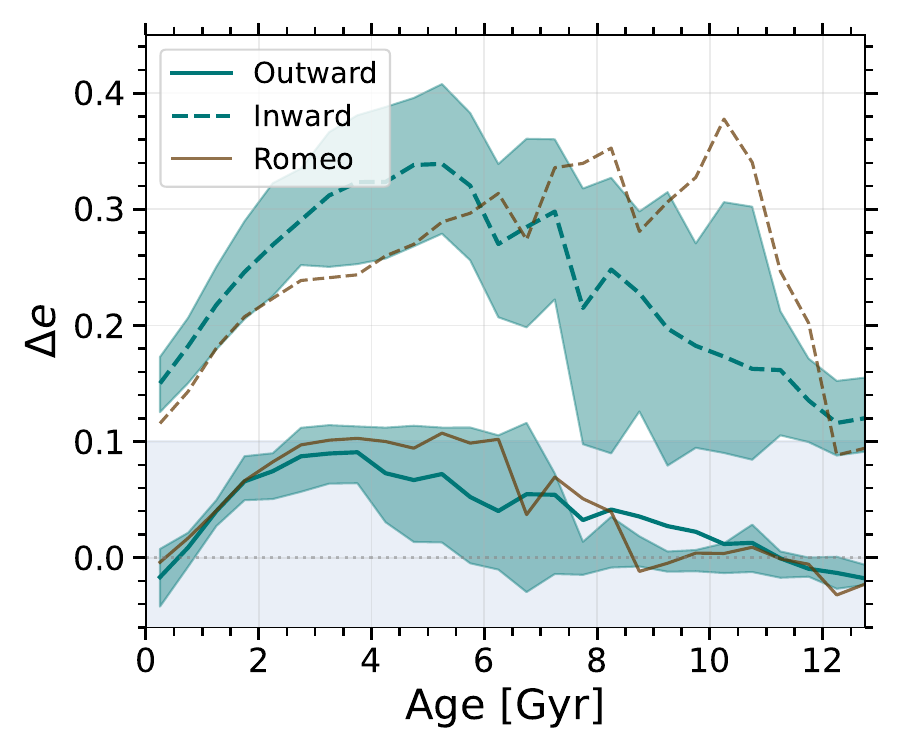}
\caption{
Median change in orbital eccentricity from birth to today, $\Delta e$, versus stellar age, for all stars (regardless of birth eccentricity) that redistributed significantly outwards (solid) and inwards (dashed).
The teal lines show the mean, shaded regions show the 68th percentile scatter across the 12 galaxies.
The blue horizontal shaded region marks stars whose dynamical evolution from birth to today was `cold', with $|\Delta e| < 0.1$.
The brown lines show Romeo, one of our earliest-forming disks, and therefore one of our best analogs to the Milky Way: it lies mostly within the galaxy-to-galaxy scatter and follows the same general trends.
In general, $\Delta e$ peaks for stars born at the transition from the early- to late-disk eras, $\approx 4 \Gyr$ ago, on average.
This is because younger stars have had less time for heating, while older stars were born too hot to experience substantial further heating.
}
\label{dele_age}
\end{figure}

Figure~\ref{dele_age} shows the median change in eccentricity, $\Delta e$, versus stellar age for inward (dashed) and outward (solid) migrators.
These results align with prior analyses of these FIRE-2 simulations \citep[for example][]{Yu2021, Yu2023}.
A maximum (average) change in eccentricity occurs around the (average) transition from the early- to the late-disk era ($\approx 4 \Gyr$ ago). In the late-disk era, the increase in $\Delta e$ with stellar age (from $0$ to $4 \Gyr$ ago) primarily reflects the amount of time stars spent undergoing dynamical scattering in a relatively settled disk. In contrast, in the early- and pre-disk eras, stars were born on intrinsically hotter orbits, leading to a saturation in subsequent heating. Despite having several additional Gyr to evolve, these populations did not experience substantially larger $\Delta e$. As a result, the intuition that dynamical heating scales monotonically with time breaks down at early epochs, where changing dynamical birth conditions play a dominant role. These cosmological simulations naturally capture these transitions, which are not accessible to idealized models that focus on dynamically settled disks.

Inward migrators show larger $\Delta e$ than outward migrators at all stellar ages. Additionally, $\Delta e$ for outward migrators remains within our fiducial definition of cold (blue shaded region) across all stellar ages, indicating that outward redistribution tends to be (moderately) dynamically cold. Importantly, while both inward and outward migrators share the same general trend with age, yet remain significantly offset in magnitude, the direction of radial redistribution is more important than age in determining $\Delta e$.

Figure~\ref{dele_age} also shows this trend for Romeo, one of our earliest-forming disks, to demonstrate that a galaxy with an early disk-onset, like the Milky Way, falls largely within the galaxy-to-galaxy scatter and follows the same general trend. Romeo's earlier disk onset at 10.25 Gyr is reflected in this figure as this is where $\Delta e$ peaks for inward migrators. However, where this line peaks is different from the average trend; this suggests that the trend of $\Delta e$ with stellar age is contingent on the formation history of the galaxies. Although the trend for Romeo peaks earlier than the average for inward-migrating stars, the trend for outward migrating stars follows the average. This is because of the inside-out radial growth of the galaxy. \cite{Graf2025b} showed that Romeo had accelerated early star formation at small radii relative to the other galaxies. Therefore, its inner regions formed earlier and reached high stellar density faster, whereas the outer regions formed more self-similarly with the other galaxies. Romeo's earlier peak for inward migrators arises because its inner disk formed earlier and more rapidly than in other galaxies, leading to enhanced earlier heating in the inner regions. The outer disk formed more typically, causing outward migrators to follow the general behavior.

\subsection{Cold-Torqued Stars}

An important expected driver of cold torquing (as described in \S\ref{sec:Introduction} and also referred to as `churning') is the exchange of angular momentum at corotation resonances of non-axisymmetric features, like transient spiral arms or bars, allowing stars to move radially without significant heating.
Many works \citep[for example][]{SellwoodBinney2002, Frankel2020, Lehmann2024} claim that this is a significant mechanism or mode of radial redistribution in disks like the MW. Because such processes produce little orbital heating, they leave weak dynamical signatures and are difficult to detect, making their overall impact on present-day stellar distributions uncertain.

In this section, we aim to quantify, phenomenologically, how prevalent cold torquing is in the disks of FIRE galaxies, between stellar birth and today.
Specifically, or each star particle, we measure its net change in $j_\phi$ and $e$ between its birth and today.
We do not track detailed orbits over time, to test the key dynamical processes at play or to investigate whether torquing and heating occur separately or at the same time throughout a star particle's history.
We are motivated by applications to galactic archeology, to quantify the link between a star's orbit today and its birth conditions.

\subsubsection{Definition of Cold-Torqued Stars}

As before, we focus on star particles that underwent significant radial redistribution in terms of changes to $j_\phi$ (torquing).
We define a star particle as significantly torqued if $|\Delta j_{\phi} / j_{\rm \phi, birth}| > 0.2$ between birth and today.
We are interested in the fraction of such star particles for which the change to the orbit is dynamically `cold', in terms of a minimal \textit{net} change to orbital eccentricity from birth to today.
A star particle has been `cold torqued' if it has significantly torqued and its $|\Delta e| < 0.1$.
Our primary metric is the cold-torqued fraction: the fraction of star particles that have cold torqued out of all stars that have (significantly) torqued.
With this, we seek to quantify how often radial redistribution (via torquing), between birth and today, is dynamically cold.

Additionally, to test the standard scenario for cold torquing, we examine specifically stars that formed on near-circular orbits ($e_{\rm birth} < 0.2$).
That said, we also explore trends when imposing no restrictions on birth $e_{\rm birth}$.
A restriction on $e_{\rm birth}$ does not matter much for stars born during the late-disk era, but as we show below, it significantly affects the trends for stars born on generally non-circular orbits during the early-disk or pre-disk eras.
Furthermore, in Appendix~\ref{sec:ct_thresh}, we explore trends for different thresholds in $|\Delta j_{\phi} / j_{\rm \phi, birth}|$ and $|\Delta e|$.

\begin{figure*}[!ht]
\centering
\includegraphics[width = 0.975 \linewidth]{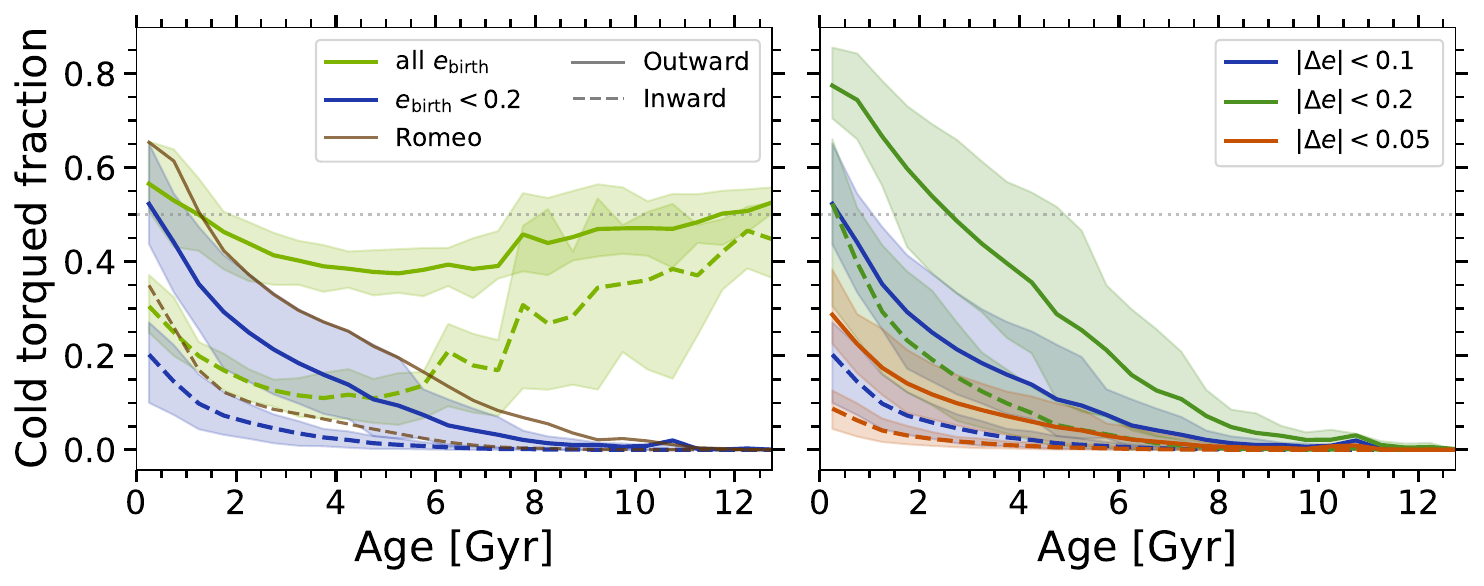}
\caption{
\textbf{Left}: Cold-torqued fraction versus stellar age, separately for outward and inward migrators.
We define the cold-torqued fraction as the number of star particles that experienced minimal changes in orbital eccentricity relative to the total number that redistributed significantly (in the given direction).
Lines show the mean, shaded regions show the 68th percentile scatter across the 12 galaxies.
We show results separately for only stars born on near-circular orbits ($e_{\rm birth} < 0.2$) and all stars.
For all but the youngest stars, the cold-torqued fraction is much higher for stars born on non-circular orbits.
Generally, radial redistribution is not cold, except for the youngest outward migrators. 
Brown lines show the trends for Romeo (using $e_{\rm birth} < 0.2$), one of our earliest-forming disks, to demonstrate that a galaxy with an early disk onset, like the Milky Way, follows the same general trends, though with a slightly higher cold-torqued fraction at a given age.
\textbf{Right}: Same, for stars with $e_{\rm birth} < 0.2$, but comparing three threshold for cold torqued: $|\Delta e| < 0.1$, 0.2, 0.05.
Our fiducial threshold is $|\Delta e| < 0.1$.
}
\label{ct_age_efs}
\end{figure*}

\subsubsection{Dependence on Age and Migration Direction}

Figure~\ref{ct_age_efs} shows the cold-torqued fraction versus stellar age, separated by migration direction.
We measure this separately for stars that migrated inward and for stars that migrated outward.
Figure~\ref{ct_age_efs} (left) shows this fraction when imposing, and when not imposing, a selection on birth eccentricity.

We focus first on stars born on near-circular orbits, $e_{\rm birth} < 0.2$, which is the most common scenario explored in the literature.
As in the previous section, there is a key difference between inward and outward migrators.
Outward migrators consistently show a significantly higher cold-torqued fraction at all ages.
We interpret that this asymmetry arises because outward migrators moved into lower-density regions where heating was less likely, while inward migrators entered denser environments where heating became increasingly unavoidable.

Furthermore, for both inward- and outward-migrating stars, the age dependence is strong.
Only for the youngest stars ($\lesssim 1 \Gyr$) that migrated outward is the cold-torqued fraction more than half, reaching up to $\approx 55\%$, on average.
For inward-moving stars, the fraction is always $\lesssim 20\%$.
Thus, for stars born cold, the direction of migration is crucial in determining whether the stars can preserve their low eccentricities.
Both fractions decline rapidly with age: older stars born on near-circular orbits rarely stay on near-circular orbits.
Therefore, for stars born on near-circular orbits, we conclude that radial redistribution is generally not cold.

Figure~\ref{ct_age_efs} shows results separately for Romeo, one of our earliest-forming disks and best MW analogs.
Romeo shows the same general trend as average, but because of its earlier-forming disk, with colder dynamics today, its cold-torqued fraction is mildly elevated by $\approx 10\%$ for both inward- and outward-migrating stars. In Section~\ref{subsec:indv_gal} we investigate this further and show these results separately for our 4 earliest-forming galaxies.

Figure~\ref{ct_age_efs} (left) also shows stars born across all eccentricities.
Now, the dependence on age is much weaker.
This reflects the changing orbits that stars were born on across cosmic time.
Most stars born during the late-disk era formed on near-circular orbits, so the results are similar, regardless of an upper limit on $e_{\rm birth}$.
However, stars born during the early-disk and pre-disk eras formed on more eccentric orbits.
Therefore, the imposition that $e_{\rm birth} < 0.2$ limits to the small fraction of such stars that were born on near-circular orbits, and because of the hotter overall dynamics of the disk at those earlier times, few such stars can remain on near-circular orbits to today.
When we do not impose a cut on $e_{\rm birth}$, the cold-torqued fraction shows only weak dependence on age.
This is because, while older stars have had more time to be heated, they also were born hotter, making them more difficult to heat further.

Indeed, Figure~\ref{ct_age_efs} (left) shows that both inward and outward migrators exhibit a broad minimum in the cold-torqued fraction at intermediate ages, $\approx 4 \Gyr$ ago.
This is similar to the peak in $\Delta e$ in Figure~\ref{dele_age}.
At late times ($\lesssim 4 \Gyr$), the disk is settled, and the subsequent dynamical heating depends primarily on how long stars have had to heat rather than on their birth conditions.
$\gtrsim 4 \Gyr$ ago, both cold-torqued fractions gradually rise with age, especially the inward migrators, whose cold-torqued fraction is comparable to the outward migrators $\sim 10 \Gyr$ ago.
The oldest stars were born dynamically hot, so moving inward versus outward makes comparatively little difference, because the stars remain similarly hot throughout their radial redistribution.

With our fiducial definition of cold ($|\Delta e| < 0.1$), Figure~\ref{ct_age_efs} (left) shows that radial redistribution is consistently not cold across all stellar ages today.
Even in the late-disk era, only up to about half of outward-migrating stars preserved birth eccentricity.
However, the cold-torqued fraction is sensitive to the threshold of $|\Delta e|$ to define cold.
Therefore, Figure~\ref{ct_age_efs} (right) shows the cold-torqued fraction for stars born on near-circular orbits ($e_{\rm birth} < 0.2$), comparing 3 definitions of `cold torqued': $|\Delta e| < 0.1$, $0.2$, and $0.05$.
Doubling the threshold to $|\Delta e| < 0.2$ increases the cold-torqued fraction, but only moderately, by $\approx 20\%$.
Now, a majority of young outward-migrating stars remain cold, but the fraction drops below 50\% within $\approx 2.5 \Gyr$.
Even with this more generous definition of `cold', less than half of inward-migrating stars cold torqued.
When restricting the threshold to $|\Delta e| < 0.05$, the fraction of young outward migrators decreases to at most $\approx 30\%$, and for inward migrators, the cold-torqued fraction only reaches $\approx 5\%$.
Regardless of the threshold in $|\Delta e|$, the age dependence is strong, such that radial redistribution between birth and today is generally not cold except for outward migrators born recently.

\subsubsection{Dependence on Eccentricity at Birth and Today}

\begin{figure*}[!ht]
\centering
\includegraphics[width = 0.975 \linewidth]{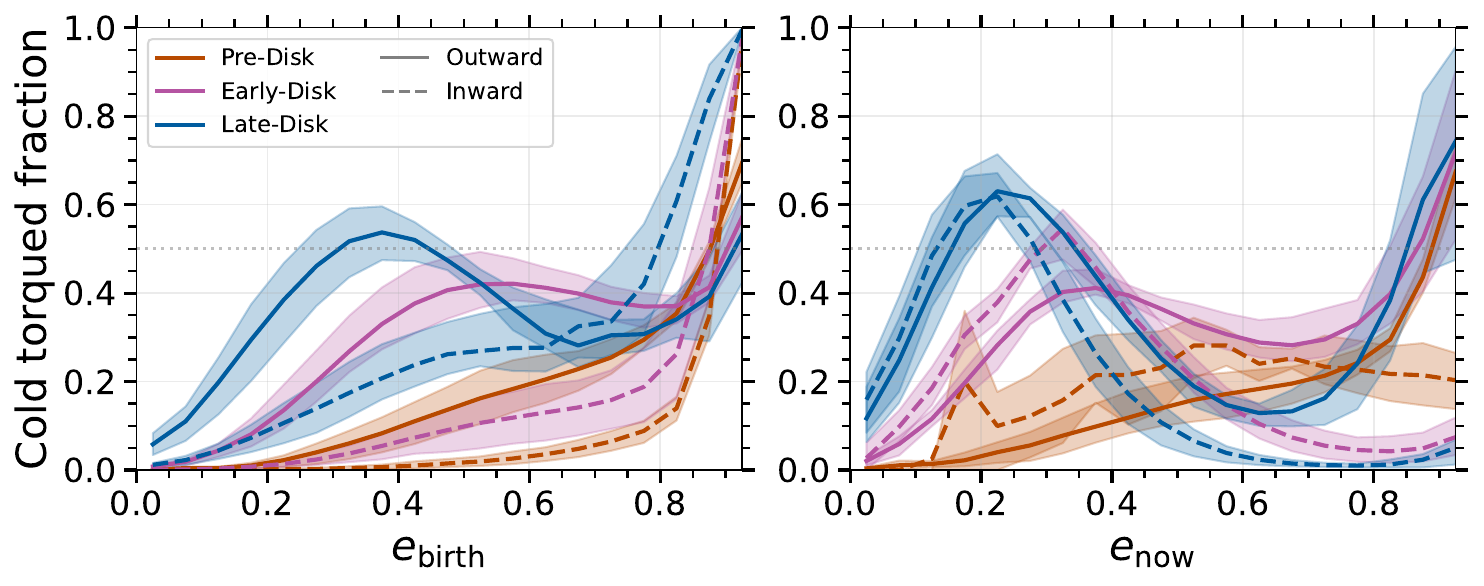}
\caption{
Cold-torqued fraction versus orbital eccentricity at birth (left) and today (right), for stars separated into the three disk eras in which they were born, and separately for inward- and outward-migrating stars.
The cold-torqued fraction is defined as the number of stars that experienced minimal change in eccentricity relative to all stars of a given eccentricity that were significantly radially redistributed inward or outward.
Lines show the mean, shaded regions show the 68th percentile scatter across the 12 galaxies.
Stars born on moderately eccentric orbits ($\approx 0.4$) are more efficient at preserving their eccentricity during significant radial redistribution than those born on circular orbits. By contrast, stars found today with ($e_{\mathrm{now}}\approx 0.2$) are also most likely to have been cold torqued. This apparent offset arises because the present-day eccentricity distribution reflects the combined effect of the birth eccentricity distribution and the broad distribution of $\Delta e$, such that the eccentricity most likely to have $\Delta e \approx 0$ does not correspond to the same value at birth and at present day.
}
\label{ct_efe}
\end{figure*}

Figure~\ref{ct_efe} shows the cold-torqued fraction binned by eccentricity at birth (left) and today (right).
Specifically, for all stars at a given eccentricity at birth (or today) that experienced significant radial redistribution (torqued), we measure the fraction with $|\Delta e| < 0.1$ between birth and today.
This addresses the question of whether stars born on more circular orbits are more likely to radially redistribute without being dynamically heated.
Figure~\ref{ct_efe} shows that this is not necessarily true.
For all three disk eras, a majority of stars born on extremely eccentric orbits radially redistribute with minimal changes in eccentricity. In general, the cold-torqued fraction increases with birth eccentricity. For inward migrators, the fraction gradually rises at low $e_\mathrm{birth}$ followed by a steeper increase toward $e_\mathrm{birth} \approx 1$. Outward migrators show a steeper rise that plateaus. For late-disk outward migrators, the fraction peaks around $e_\mathrm{birth} \approx 0.4$ and then saturates. Early-disk stars show a similar pattern, with outward migrators peaking near $e_\mathrm{birth} \approx 0.5$ before leveling off. Pre-disk outward migrators instead show a steady increase in the cold-torqued fraction with birth eccentricity up to 70\%.

The cold-torqued fraction is highest for outward-migrating late- and early-disk stars born with eccentricities $e_\mathrm{birth} \approx 0.3 - 0.5$. One contribution to this dependence is that stars born on more circular orbits are more likely to form and stay near the disk midplane, so they may be exposed to stronger perturbations than stars on moderately eccentric orbits. Furthermore, stars born on moderately eccentric orbits that experience multiple scattering events throughout their redistribution can have their eccentricities evolve in both directions (heating or cooling), such that the net change remains small and their present-day eccentricities remain similar to their birth values. Although late-disk stars with $e_{\rm birth} \approx 0.4$ are the most likely to preserve eccentricity, the bulk of this population today was born colder. The jump to a cold-torqued fraction of $\approx 60\%$ at $e_{\rm birth} \approx 0.9$ for the late-disk era consists of rare stars. They likely represent extreme scattering events that, like pre-disk stars, are born at the ceiling of their phase space and cannot be heated further.

Together, these trends demonstrate that stars can radially redistribute with minimal changes in eccentricity across the full range of birth eccentricities, which challenges the common assumption that radial redistribution necessarily leads to dynamical heating and that preservation of eccentricity throughout redistribution applies to only those on initially circular orbits (`provenance bias').

Figure~\ref{ct_efe} (right) shows the cold-torqued fraction binned by present-day eccentricity, which highlights a somewhat different picture. The distinction between inward and outward migrators is much smaller at low $e_{\rm now}$ and becomes more pronounced at high $e_{\rm now}$, similar to the behavior in Figure~\ref{dele_efe} (right). For late-disk stars, the cold-torqued fraction rises up to $e_{\rm now} \approx 0.2$. At higher $e_{\rm now}$, the inward-migrating population declines to zero, while outward migrators continue to show substantial cold-torqued fractions up to $\approx 70\%$. (Late-disk stars today with $e_{\rm now} > 0.6$ are rare.) Similarly, the fraction for early-disk stars increases up to $e_{\rm now} \approx 0.4$, after which the inward migrators fall to zero, and the outward migrators rise toward a fraction of $\approx 70\%$. Pre-disk stars, in contrast, show inward migrators maintaining a nearly constant cold-torqued fraction of $\approx 20\%$ across the full $e_{\rm now}$ range and outward migrators steadily increasing toward a fraction of $\approx 70\%$ at high $e_{\rm now}$.

Figure~\ref{ct_efe} reinforces the idea that present-day eccentricity does not simply trace birth eccentricity, in terms of being cold torqued.
When selecting stars based on $e_{\rm now}$, particularly at low eccentricities, it does not matter much whether a star migrated inward or outward. Additionally, near-circularity at birth does not improve the preservation of an orbit's circularity during radial redistribution. Instead, Figure~\ref{ct_efe} demonstrates that stars born on moderately eccentric orbits are more likely to preserve their birth eccentricities after significant radial redistribution than stars born on near-circular orbits, and that outward migration is the dominant channel for maintaining orbital eccentricity, regardless of the era in which a star was born.

\subsubsection{Dependence on Radius, at Birth and Today}

\begin{figure*}[!ht]
\centering
\includegraphics[width = 0.975 \linewidth]{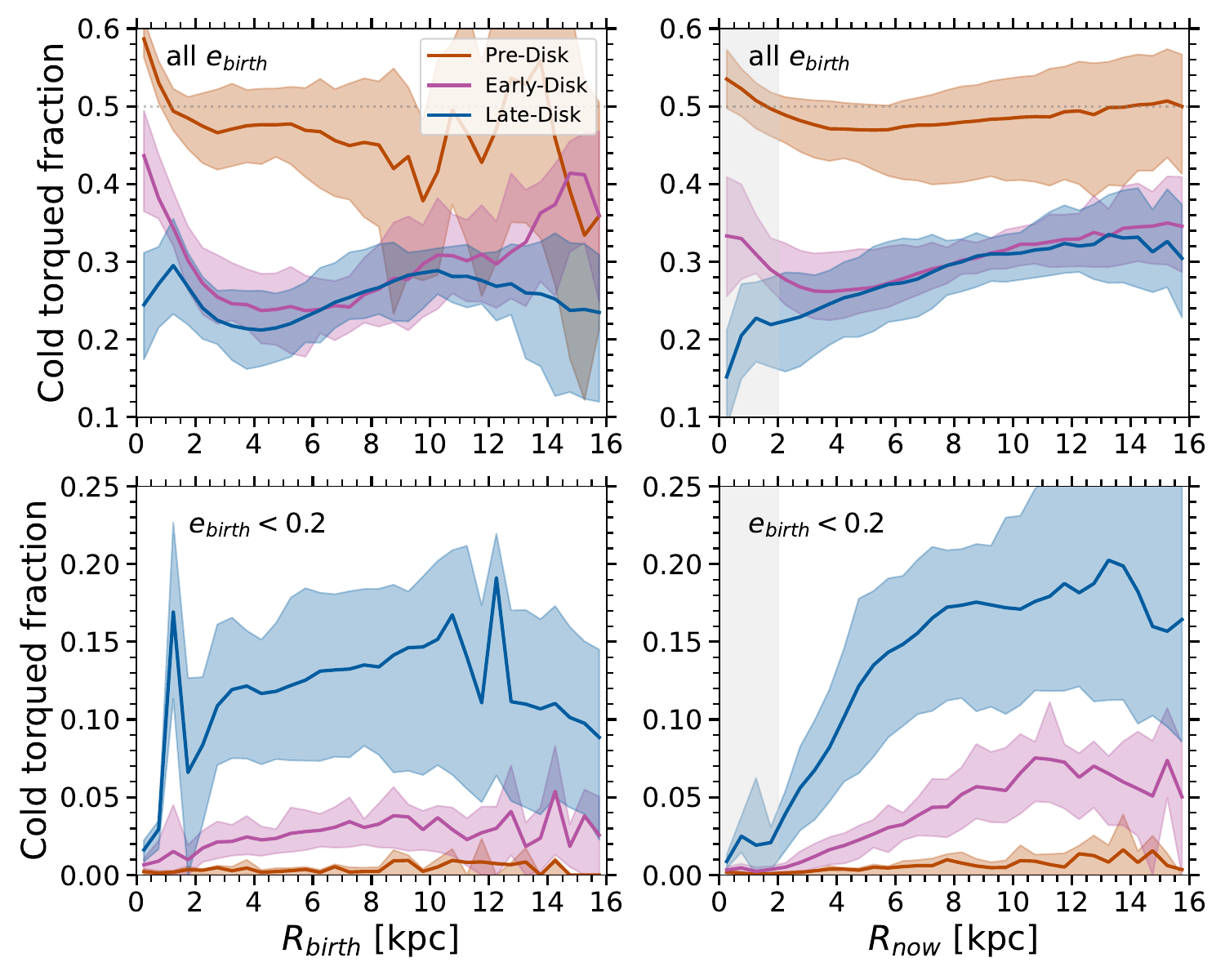}
\caption{
Cold-torqued fraction versus radius at birth (left) and today (right), for stars separated into the three disk eras in which they were born.
The top row includes all stars, while the bottom row includes only stars formed on near-circular orbits ($e_{\rm birth} < 0.2$).
The cold-torqued fraction is defined as the fraction of stars that undergo significant radial redistribution while experiencing only minimal changes in eccentricity. 
Lines show the mean, shaded regions show the 68th percentile scatter across the 12 galaxies.
The shaded region at $R_{\rm now} = 0 – 2 \kpc$ denotes the inner galaxy, which we exclude from the main analysis.
Across all eras and eccentricities, the cold-torqued fraction is largely independent of radius outside the inner galaxy, indicating that being cold torqued is not preferentially associated with a particular birth radius or final radius. The strongest radial dependence appears in the $R_{\rm now}$ panel for stars born on near-circular orbits, where the cold-torqued fraction declines sharply toward small radii ($<5 \kpc$ for stars born during the late-disk era) and remains flat otherwise.
}
\label{ct_r_efs_4pan}
\end{figure*}

Figure~\ref{ct_r_efs_4pan} shows the cold-torqued fraction versus birth radius (left column) and present-day radius (right column), for all stars regardless of birth eccentricity (top) and for stars formed on near-circular orbits ($e_{\mathrm{birth}} < 0.2$; bottom), separated into the three disk eras. Examining trends with radius allows us to test whether being cold torqued is preferentially associated with stars born at specific radii or ending up at particular radii today.

Across all eras and for both eccentricity selections, the cold-torqued fraction is relatively flat with radius once the inner galaxy region is excluded. We shade the region $R_{\mathrm{now}} = 0 - 2 \kpc$ to indicate the stars currently in the inner galaxy that we exclude from the majority of our analysis.

Figure~\ref{ct_r_efs_4pan} (top row) shows results consistent with previous sections. Pre-disk stars are the most efficient at preserving their eccentricities through significant radial redistribution, with a nearly constant cold-torqued fraction of $\approx 50\%$ across all birth radii and present-day radii. The early- and late-disk populations follow similar trends to each other, overlapping across most radii with a steady fraction of $\approx 30\%$. For stars born within the inner $2 \kpc$, all eras show a slight increase in the cold-torqued fraction relative to larger radii. This reflects the fact that these stars are born in the dynamically hot inner galaxy, where the bulge and bar exist, and, as shown previously, stars that are born hot are more likely to remain hot. Appendix~\ref{sec:radial_e} shows that stars born in the inner $2 \kpc$ on average have higher birth eccentricities, and also on average experience the lowest changes in eccentricity from birth to today. Conversely, when considering $R_{\mathrm{now}}$, the early- and late-disk populations show a slight downturn within the inner $2 \kpc$, because stars that migrate into this dynamically hot region are more easily heated from moderate or initially cold eccentricities.

Figure~\ref{ct_r_efs_4pan} (bottom row) includes only stars born on near-circular orbits.
Here, late-disk stars are the most likely to preserve their eccentricities across all birth radii and present-day radii, followed by the early- and then pre-disk populations. We expected this behavior because late-disk stars predominantly formed on circular orbits, early-disk stars formed on moderately eccentric orbits, and almost no pre-disk stars formed on circular orbits, resulting in no pre-disk stars being cold torqued across all radii. Excluding the inner galaxy, late-disk stars exhibit an average cold-torqued fraction of $\approx 12\%$ with $R_{\mathrm{birth}}$ and $\approx 17\%$ with $R_{\mathrm{now}}$.

The sharp increase in the cold-torqued fraction for late-disk stars born within the inner $2 \kpc$ is a result of rare stars. As Appendix~\ref{sec:radial_e} shows, the average birth eccentricity of late-disk stars in the inner galaxy is $\approx 0.6$, meaning that few stars satisfy $e_{\mathrm{birth}} < 0.2$ in this region. In contrast, when examining the cold-torqued fraction with $R_{\mathrm{now}}$, the fraction for late-disk stars decreases sharply from $\approx 15\%$ to $\approx 1\%$ within the inner $5 \kpc$. 
On average, late-disk outward migrators are born at $\approx 5\,\mathrm{kpc}$, while inward migrators are born at $\approx 7\,\mathrm{kpc}$. The sharp decline in the cold-torqued fraction at $R_{\mathrm{now}} \lesssim 5\,\mathrm{kpc}$ indicates that the inner galaxy is dominated by inward migrators, for which $\Delta e$ is systematically larger. 
The fraction reaches $\approx 1\%$ within the inner $2 \kpc$, further implying that stars born on circular orbits have essentially no chance of preserving their circularity if they redistributed into the inner galaxy.

Figure~\ref{ct_r_efs_4pan} reinforces that the distinction between the three dynamical eras remains consistent regardless of location and that the cold-torqued fraction is primarily driven by a star's dynamical birth conditions rather than its specific position in the disk. While factors like stellar age and migration direction play a large role in the cold-torqued fraction, the radial trends in Figure~\ref{ct_r_efs_4pan} are relatively flat across the majority of the disk. This suggests that the physical processes governing whether a star preserves its eccentricity are widespread properties of the disk during a given dynamical era rather than being localized to specific radii.

\subsubsection{Dependence on Galaxy Properties}
\label{subsec:indv_gal}

\begin{figure*}[!ht]
\centering
\includegraphics[width = 0.975 \linewidth]{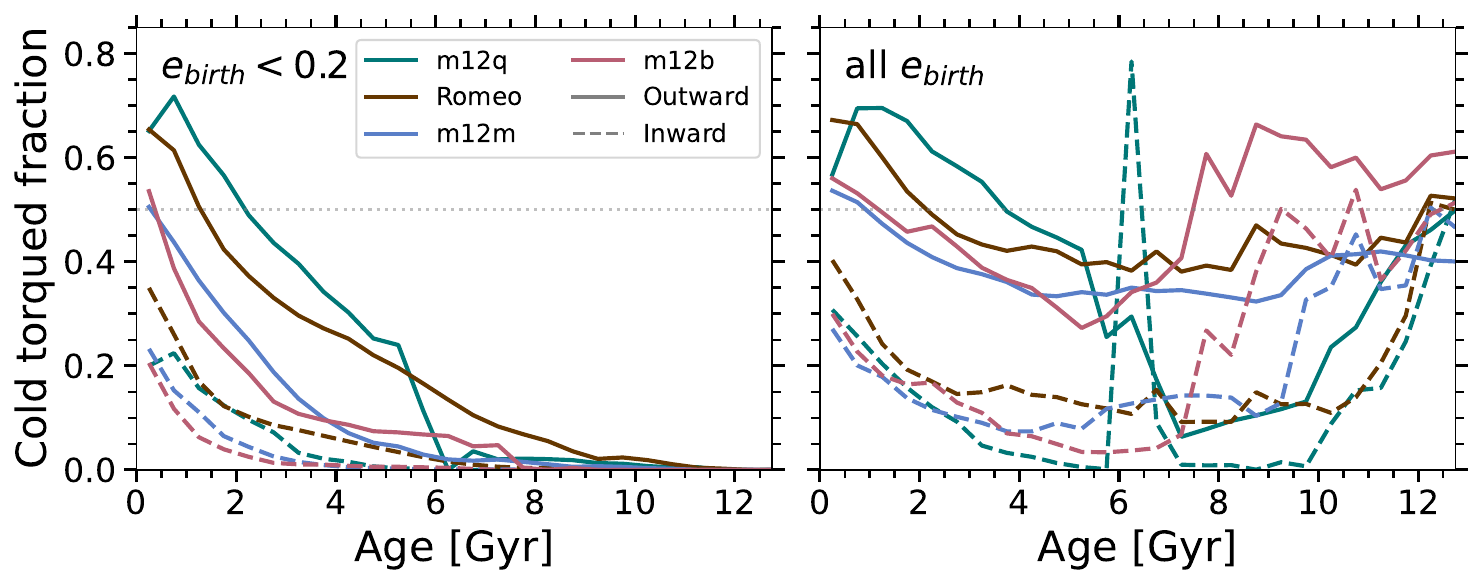}
\caption{
\textbf{Left}: Cold-torqued fraction versus stellar age for stars with $e_{\rm birth} < 0.2$ and separated by migration direction. We show our four earliest-forming galaxies, which also have the highest $v_{\phi} / \sigma_{\rm v}$ today, ordered from earliest to latest disk onset: m12q, Romeo, m12m, and m12b. The cold-torqued fraction is highest in m12q, which has the earliest disk onset, and decreases with later disk onset times. \textbf{Right}: The same, but for all stars regardless of $e_{\rm birth}$.
}
\label{ct_vs_age_m12s}
\end{figure*}

The cold-torqued fraction is highest among the youngest stars and in the earliest-forming disks, suggesting a possible connection between the prevalence of being cold torqued and the formation history of the galaxy. Here, we examine this relationship more directly by comparing the galaxies individually. Figure~\ref{ct_vs_age_m12s} shows the cold-torqued fraction with age for inward- and outward-migrating stars with $e_{\rm birth} < 0.2$ (left panel) and all stars regardless of  $e_{\rm birth}$ (right panel). To investigate whether our general results are consistent with an early disk onset, like the Milky Way, we show these trends for our four earliest-forming galaxies, ordered from earliest to latest disk onset: m12q, Romeo, m12m, and m12b. For stars born on near-circular orbits, the cold-torqued fraction is highest in the galaxy with the earliest disk onset (m12q) and decreases with progressively later disk onset times. Therefore, the likelihood of being cold torqued increases in earlier-forming disks.
We also examined these trends for stars within the solar annulus ($8 \pm 2 \kpc$), and we find the same overall trend and ordering of galaxies. Including all stars regardless of birth eccentricity (right panel), the rank ordering of galaxies becomes noisier, especially at early times.

\begin{figure*}[!ht]
\centering
\includegraphics[width = 0.975 \linewidth]{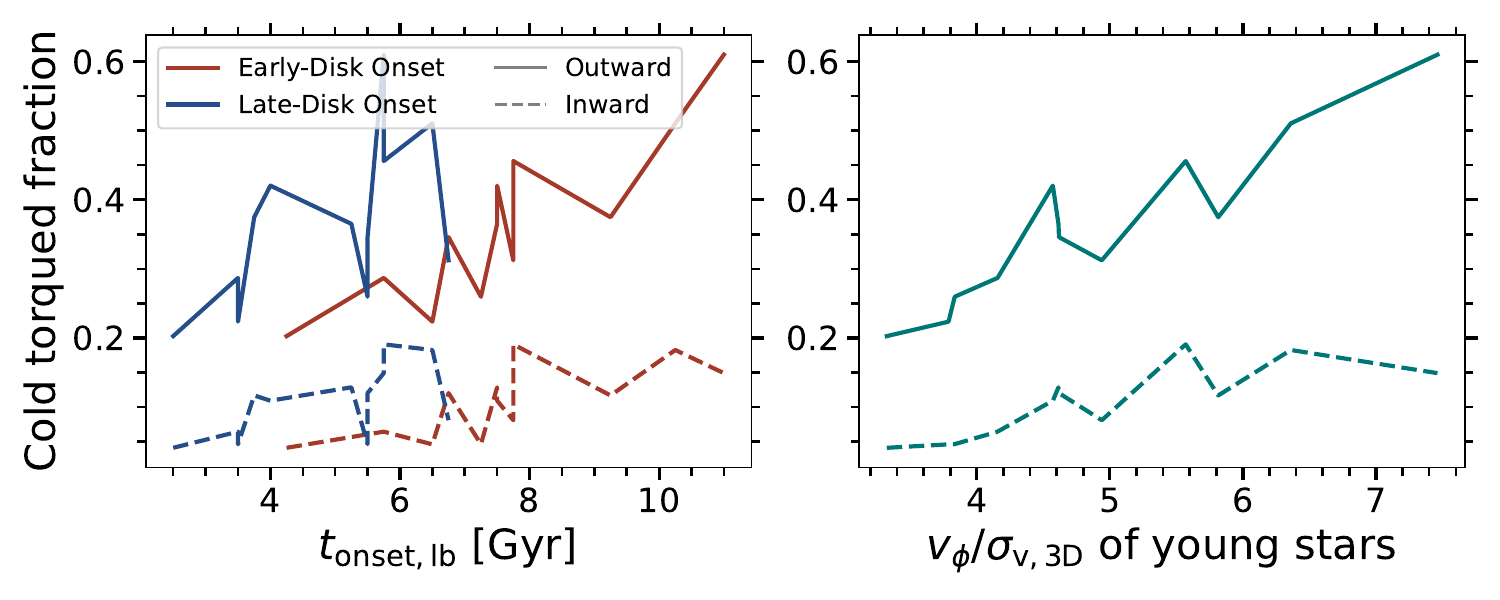}
\caption{
\textbf{Left}: Cold-torqued fraction of stars with ages $< 2 \Gyr$ and $e_{\rm birth} < 0.2$, shown separately for inward and outward migrators in each galaxy, versus the lookback time of the galaxy’s early disk onset (when $v_{\phi} / \sigma_{v, \rm 3D} > 1$) and late disk onset (when $v_{\phi} / \sigma_{v, \rm 3D} > 3$).
\textbf{Right}: The same, versus the present-day $v_{\phi} / \sigma_{v, \rm 3D}$ measured for stars with ages $< 100 \Myr$, a metric for the galaxy's current diskiness. Galaxies with earlier-forming disks, and/or dynamically colder disks today, show higher cold-torqued fractions for both inward- and outward-migrating stars. Given that the Milky Way is thought to be an early-forming, dynamically cold disk, this implies that it may exhibit an enhanced cold-torqued fraction, more consistent with our most extreme galaxies in FIRE.
}
\label{ct_onset_vsig}
\end{figure*}

Figure~\ref{ct_onset_vsig} shows the cold-torqued fraction for stars born in the last $2 \Gyr$ on near-circular orbits ($e_{\mathrm{birth}} < 0.2$) versus the galaxy's early-disk onset  (when $v_{\phi} / \sigma_{v, \rm 3D} > 1$) and late-disk onset (when $v_{\phi} / \sigma_{v, \rm 3D} > 3$) lookback times (left panel) and the median $v_{\phi} / \sigma_{v, \rm 3D}$ for stars $< 100 \Myr$ old (right panel), allowing us to assess whether earlier-forming and more rotation-supported disks systematically exhibit higher cold-torqued fractions among their youngest stellar populations.

Figure~\ref{ct_onset_vsig} (left) shows that the cold-torqued fraction for outward migrators increases in galaxies with earlier disk onset times, rising from $\approx 20\%$ in the latest-forming disk ($t_{\rm onset} = 4.25 \Gyr$ ago) to $\approx 60\%$ in the earliest-forming disk ($t_{\rm onset} = 11 \Gyr$ ago).
Interestingly, the trend is somewhat weaker/noisier for the late-disk onset time, with fractions generally increasing toward earlier late-disk onsets.
Similarly, the cold-torqued fraction for inward migrators increases with disk onset time. \cite{Bellardini2026} found that the total amount of radial redistribution across $R = 2 - 12 \kpc$ does not correlate strongly with disk onset time, regardless of the redistribution metric used. Combined with our results, this suggests that earlier-forming disks do not necessarily experience more radial redistribution overall. Instead, they experience more redistribution that is cold.

\cite{McCluskey2024} showed for these same FIRE galaxies that earlier-forming disks tend to be dynamically colder today. Consistent with this, Figure~\ref{ct_onset_vsig} (right) shows that the cold-torqued fraction increases with a galaxy's $v_{\phi} / \sigma_{v, \rm 3D}$ of young stars today. In fact, the correlation between the cold-torqued fraction and $v_{\phi} / \sigma_{v, \rm 3D}$ today is stronger than those with the disk onset times. Together, these results indicate that the dynamical state of the disk, closely tied to its formation history, affects how commonly radial redistribution (via torquing) is cold.

We also examined these trends measuring the cold-torqued fraction for stars born throughout the late-disk era (instead of over a fixed age range of $2 \Gyr$). The fractions follow the same qualitative behavior, but the correlations are weaker. This makes sense, because earlier-forming disks have longer late-disk eras, leading to a longer time interval over which dynamical heating can occur, resulting in lower and flatter cold-torqued fractions across an entire late-disk era, as compared with a fixed age range. Appendix~\ref{sec:corr_coeff} shows Pearson and Spearman correlation tests for these results, quantifying the difference between selecting stars across age $< 2 \Gyr$ and across the late-disk era.

The MW is an unusually early-forming galaxy, and \cite{McCluskey2025} showed that it is also a kinematic outlier, with significantly colder dynamics than both other observed galaxies and the FIRE-2 simulations. Given our finding that colder, earlier-forming disks exhibit higher cold-torqued fractions, this suggests that stars being cold torqued may be more common in the MW than in our simulated galaxies and more common than in a typical disk.

\subsubsection{Summary of the Main Dependencies}

Table~\ref{tab:parameter_importance} ranks the parameters we explored by the strength of their correlation with changes to eccentricity and the cold-torqued fraction. Although stellar age correlates strongly with birth eccentricity, we measure their effects separately and find that birth eccentricity plays a more fundamental role in determining whether stars are cold torqued. In particular, stars born maximally circular are the least likely to remain cold.

\begin{table*}
\centering
\caption{
Relative importance of key parameters on changes to stellar orbital eccentricity and being cold torqued
}
\label{tab:parameter_importance}
\begin{tabular}{|c|ccccc|}
\hline
Rank & Parameter &
$\Delta e$ (inward) & $\Delta e$ (outward) &
cold-torqued fraction (inward) & cold-torqued fraction (outward) \\
\hline

1 & migration direction &
strong &
strong &
strong &
strong
\\

2 & birth eccentricity $e_{\mathrm{birth}}$ &
strong 
& strong 
& strong 
& strong 
\\

3 & stellar age &
strong 
& weak 
& strong 
& strong 
\\

4 & disk thinness $v_{\phi} / \sigma_{v, \rm 3D}$ 
& weak  
& weak 
& moderate 
& strong 
\\

5 & redistribution amount $\Delta j_\phi / |j_{\phi,\rm birth}|$ &
strong 
& moderate 
& strong 
& weak 
\\

6 & birth radius $R_{\rm birth}$ &
weak 
& weak 
& weak 
& weak 
\\
\hline
\end{tabular}
\tablecomments{
We rank the relative importance of each parameter based on the maximum range of variation, $\Delta$, in the relevant quantity. We classify effects as strong for $\Delta > 0.2$, moderate for $\Delta = 0.1 - 0.2$, and weak for $\Delta < 0.1$. Unless we vary birth eccentricity explicitly, we compute trends for stars born on near-circular orbits ($e_{\rm birth} < 0.2$). For trends with stellar age, we do not impose age selection; otherwise, we measure stars born in the late-disk era of each galaxy. We separate all values by migration direction (inward versus outward), because it is the dominant parameter.
}
\end{table*}

\subsection{Dynamical Cooling}

As Figure~\ref{dele_efen_dists} showed, most stars were not born on perfectly circular orbits, and as Figure~\ref{dele_efe} showed, their orbits can become significantly more circular over time.
We refer to these stars as having `dynamically cooled'.
To define such a population, we adopt a threshold of $\Delta e < -0.1$, to ensure that the orbits have meaningfully become more circular since birth, and to ensure that this population is distinct from our definition of `cold-torqued' stars.
Also, in this subsection, we impose no threshold on birth eccentricity.

\subsubsection{Dependence on Age and Migration Direction}

\begin{figure}[ht!]
\includegraphics[width = 0.975 \linewidth]{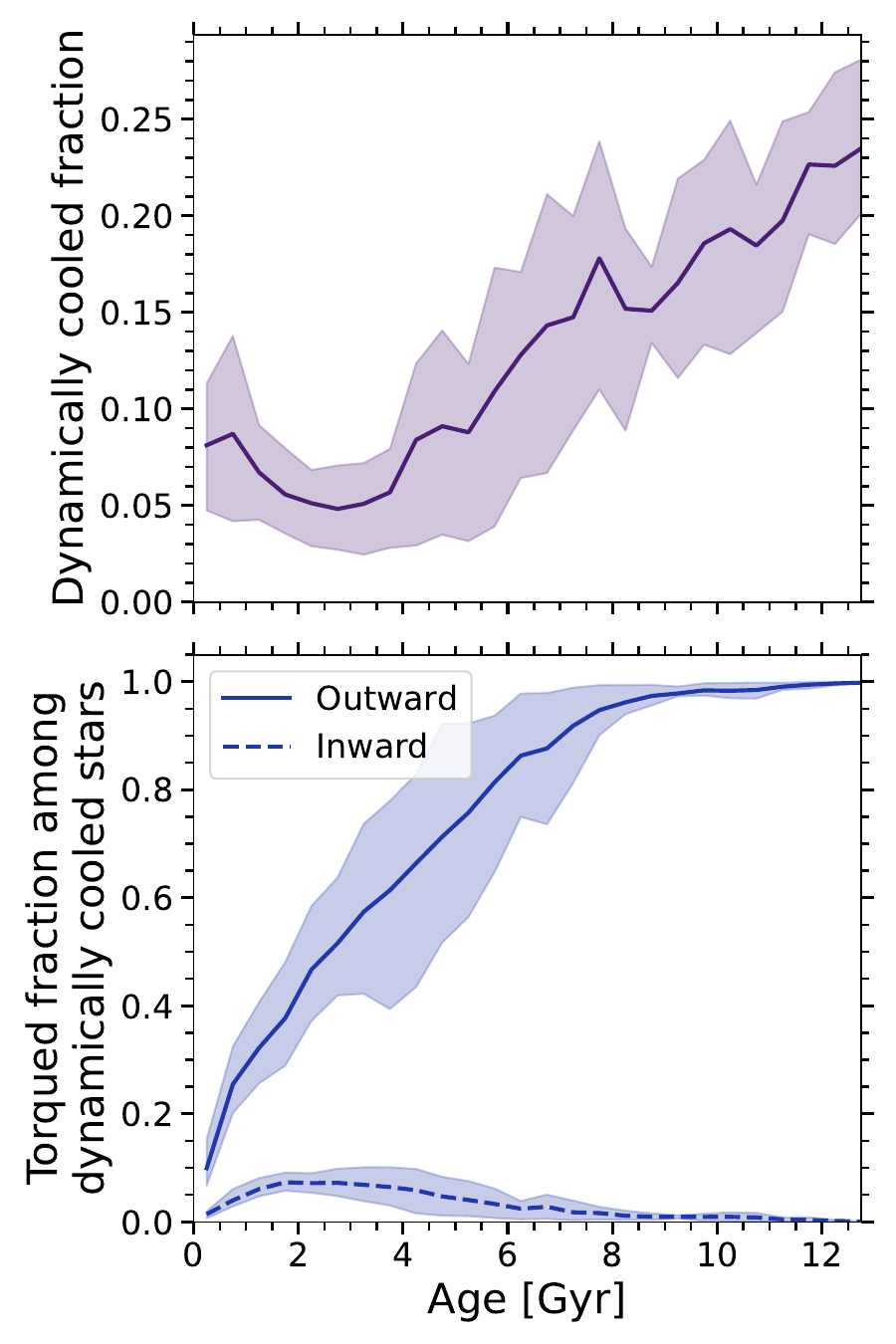}
\caption{
\textbf{Top}: Dynamically cooled fraction versus stellar age.
The fraction represents the number of stars that experienced $\Delta e < -0.1$ relative to all stars.
\textbf{Bottom}: Fraction of dynamically cooled stars that significantly torqued ($|\Delta j_\phi / j_{\phi,\rm birth}| > 0.2$), separated by migration direction, versus stellar age.
Solid and dashed lines correspond to outward and inward migrators, respectively.
Here, we impose no cut on birth eccentricity.
Lines show the mean, and shaded regions show the 68th percentile scatter across the 12 galaxies.
The dynamically cooled fraction increases with stellar age, reaching up to $\approx 23\%$ for the oldest stars and remaining below $\approx 10\%$ for the young stars.
Among dynamically cooled stars older than $\approx 2.5 \Gyr$, most were also significantly torqued outward.
Toward younger ages, however, the torqued fraction among dynamically cooled stars decreases, indicating that torquing is not the dominant channel for all dynamically cooled stars.
}
\label{crt_vs_age_dele}
\end{figure}

Figure~\ref{crt_vs_age_dele} shows the dynamically cooled fraction (top) and the torqued fraction among dynamically cooled stars (bottom) versus stellar age. The dynamically cooled fraction represents the number of stars that cooled significantly ($\Delta e < -0.1$) relative to all stars. The dynamically cooled fraction increases with stellar age, reaching $\approx 23\%$ for the oldest stars, while being $\lesssim 10\%$ for young stars.
Thus, while it is only a subset of all stars, the dynamically cooled population represents a significant subset.

The bottom panel addresses the question: among stars that dynamically cooled, what fraction also significantly torqued since birth? We separate this fraction by migration direction, where being significantly torqued corresponds to $|\Delta j_\phi / j_{\phi, \rm birth}| > 0.2$. Among dynamically cooled stars older than $\approx 2.5 \Gyr$, most of them redistributed outward.
Thus, for older stars, being significantly torqued outwards is strongly associated with, and possibly leads to, dynamical cooling.
However, at younger ages the torqued fraction among dynamically cooled stars is small, so most young stars that dynamically cooled did not radially redistribute much since birth.
The torqued fraction for inward migrators is $\lesssim 10\%$ across all ages, consistent with our previous results that stars migrating significantly inward are heated at all ages.

\subsubsection{Dependence on Eccentricity Today}

\begin{figure}[!ht]
\includegraphics[width = 0.975 \linewidth]{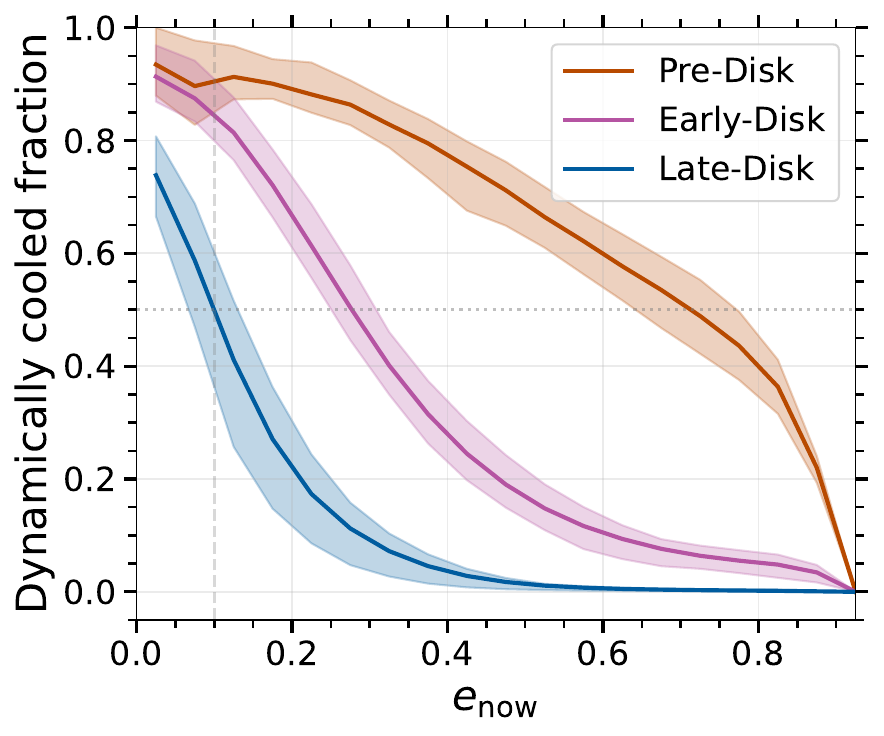}
\caption{
Dynamically cooled fraction versus orbital eccentricity today. This fraction is the number of stars that experienced a significant reduction in eccentricity ($\Delta e < -0.1$) relative to all stars born in the given era.
Lines show the mean, shaded regions show the 68th percentile scatter across the 12 galaxies.
Most stars on circular orbits today ($e_{\rm now} \lesssim 0.1$) dynamically cooled (became more circular) since birth.
}
\label{crt_efen}
\end{figure}

Figure~\ref{crt_efen} shows the dynamically cooled fraction binned by eccentricity today. The fraction increases at smaller $e_{\rm now}$.
This makes sense, given that, as we showed, stars in FIRE are generally born with some non-zero $e_{\rm birth}$, so at progressively lower $e_{\rm now}$, dynamically cooling is a progressively more important route to get to such a small $e_{\rm now}$.
This trend persists for stars born across all three disk eras, with pre-disk stars showing the highest fractions at fixed $e_{\rm now}$, followed by early- and then late-disk stars.
This dependence on disk era makes sense, because stars born earlier would have needed to dynamically cool more (on average) to reach a given $e_{\rm now}$ today.
Up to $\approx 75\%$ of late-disk and up to $\approx 90\%$ of pre- and early-disk stars now on circular orbits were born more eccentric and dynamically cooled significantly.

Previously, we defined near-circular orbits using a threshold of $e_{\rm birth} < 0.2$, consistent with the typical birth eccentricities of late-disk stars and therefore appropriate for our fiducial cold-torqued analysis. However, here we are specifically interested in the origins of the most circular orbits today, which we classify as $e_{\rm now} < 0.1$.
Across all three disk eras, the majority of stars on such circular orbits today became significantly more circular since birth.
\textit{Specifically, among all stars today on circular orbits ($e_{\rm now} < 0.1$), most (63\%) dynamically cooled, that is, they were born on more eccentric orbits and experienced a significant net decrease in eccentricity of $\Delta e < -0.1$.}
Furthermore, among all stars on circular orbits today that dynamically cooled, 38\% significantly torqued (almost always outward), suggesting that outward radial redistribution plays an important (but not always dominant) role in creating the population of stars on the most circular orbits today.
\textit{Altogether, if a star is on a circular orbit today, most likely, it did not form on such a circular orbit, but rather, it dynamically cooled.}

\section{Conclusions and Summary}

We used the FIRE cosmological simulations of Milky Way–mass galaxies to test the degree to which the radial redistribution (torquing) of stellar orbits is hot or cold.
We examined changes in orbital eccentricity from birth to today, across three dynamical eras of cosmological disk evolution.
One of our goals is to test a classic cold-torquing scenario, whereby stars born on near-circular orbits undergo specific resonant scattering processes that cause significant radial redistribution while keeping their orbits dynamically cold (near-circular) today, in the context of cosmological galaxy formation.
We examined only changes from birth to today; we did not examine full orbital histories, so our results do not necessarily speak to the prevalence of cold torquing as a process during a star's history, but we quantify the net effect on a star's orbit between birth and today.
We generally focused on stars that underwent significant radial redistribution in terms of being torqued ($|\Delta j_\phi / j_{\phi,\rm birth}| > 0.2$).
We examined both stars that were born cold, $e_{\rm birth} < 0.2$, and all stars (regardless of birth orbit) to assess how strongly this assumption shapes our results.
We summarize our primary findings below.

\begin{itemize}
    \item \textbf{Radial redistribution (torquing) is generally not cold between birth and today}

    Radial redistribution, between birth and today, is rarely cold ($|\Delta e| < 0.1$). Even in the dynamically cool late-disk era, no more than about half of outward-migrating stars preserved their birth eccentricity, with significantly smaller fractions for inward-migrating stars. The cold-torqued fraction for stars near-circular at birth is progressively smaller at older ages. However, when examining all stars, regardless of $e_{\rm birth}$, the age dependence is much weaker. 
    
    \item \textbf{Fundamental inward-outward asymmetry}

    We found a strong asymmetry between stars that redistributed outward versus inward.
    This is the strongest determinant of dynamical heating.
    Inward migration almost always leads to dynamical heating ($\Delta e > 0$), whereas almost all cold(er)-torqued stars migrated outward.

    We also explored trends with birth and present-day radius: they are generally weak. Thus, whether a star moves inward or outward is far more important for its dynamical evolution from birth to today than the radius at which it formed or currently resides.

    Contrary to common intuition, stars that did not experience significant radial redistribution are not the ones most likely to preserve eccentricity from birth to today. In fact, these stars are more likely to experience moderate heating, whereas the stars most likely to stay cold are those that moved outward significantly, with $\Delta j_\phi / |j_{\phi, \rm birth}| \approx 1$.

    \item \textbf{Significance of the disk dynamical era at birth} 

    Initial conditions set the baseline for a star's subsequent evolution. While the direction of radial redistribution is the strongest predictor of whether a star heated, cooled, or remained dynamically unchanged, birth eccentricity is the next most important factor, because it determines a star's susceptibility to subsequent heating or cooling. Late-disk stars typically formed on near-circular orbits (typical $e_{\rm birth} \approx 0.2$) and on average experienced moderate dynamical heating (typical $\Delta e \approx 0.2$), whereas pre-disk stars were born on highly eccentric orbits (typical $e_{\rm birth} \approx 1$) and remained dynamically hot, given a phase-space ceiling on further heating.

    \item \textbf{Stars born with moderate eccentricity are most likely to be cold torqued}

    Contrary to common assumptions, stars born on moderately eccentric orbits ($e_{\rm birth} \approx 0.3 - 0.5$) are the most likely to preserve their eccentricities from birth to today while undergoing significant radial redistribution (torquing).

    \item \textbf{Most stars on near-circular orbits today dynamically cooled}

    Most stars that are currently on near-circular orbits ($e_{\rm now} < 0.1$) were not born on such near-circular orbits. These stars achieved their present-day low eccentricity by dynamically cooling, often during outward migration.
    Specifically, of all stars currently on near-circular orbits, 63\% of them dynamically cooled significantly ($\Delta e < -0.1$) between birth and today. Thus, dynamical cooling is the dominant way that stars end up on near-circular orbits today.

    \item \textbf{Consistency across galaxy formation histories}

    These trends persist across galaxies with diverse disk formation histories. In particular, one of our earliest-forming disks (Romeo) falls within the overall scatter in the averaged trends, indicating that the inward–outward asymmetry and age dependence are not driven solely by later disk onsets. However, the normalization of the cold-torqued fraction varies systematically with disk onset time and dynamical state today. The earliest-forming galaxies exhibit the highest cold-torqued fractions from birth to today, while galaxies with later disk onset times show progressively lower fractions. In the earliest-forming disks, radial redistribution is most likely to be cold among the youngest ($\lesssim 2 \Gyr$) stars. Similarly, the cold-torqued fraction is highest for galaxies with dynamically colder (higher $v_{\rm \phi} / \sigma_{\rm v}$) disks. This suggests that the disk formation history influences how likely a star is to be cold torqued from birth to today, even though the general trends with age and direction of radial redistribution remain the same. That said, even for Romeo, the radial redistribution is likely to be cold only for stars younger than 2 Gyr that migrated outward.

    Finally, we note the competing roles of eccentricity on the likelihood of a star being cold torqued. On one hand, stars born on moderately more eccentric (less circular) orbits are the most likely to preserve their eccentricity from birth to today if they radially redistributed. On the other hand, dynamically colder (more circular) disks overall exhibit a higher cold-torqued fraction among their stars. Thus, one must distinguish the dynamical state of the disk as a whole from the birth eccentricity of an individual star when assessing the prevalence of being cold torqued from birth to today. 
\end{itemize}

\section{Discussion}

\subsection{Caveats}

We analyzed only the net change in orbit between a star's birth and today, without following the evolution of these orbits.
Thus, some stars we classify as cold torqued may have experienced multiple heating and cooling events that left little net change in eccentricity.
Conversely, some stars that we classify as dynamically heated first may have experienced cold torquing and then heated.

A plausible interpretation of the strong asymmetry between inward and outward migrators is that inward migrators encounter regions of higher density relative to where they were born, leading to a higher likelihood of dynamical heating even for those that move slightly inward. Outward migrators move into progressively lower-density environments where additional heating is comparatively less likely. However, we did not test this rigorously: a more definitive test would require tracking orbits of individual star particles and relating changes in eccentricity to the density of their environments along their migration paths, which we defer to future work. Recently, \cite{wiggins2025} found tentative evidence for this in the case of star clusters in these FIRE-2 simulations.

Throughout this work, we neglect the vertical components of stellar orbits. Several works examined radial redistribution in the context of vertical velocity dispersion and showed that radial redistribution primarily affects populations with smaller vertical excursions \citep[for example][]{VeraCiro2014, VeraCiro2016, Mikkola2020}. Eccentricity at birth correlates with vertical velocity dispersion, so some of our trends identified partially may reflect reduced interaction with the thin disk for stars on dynamically hotter orbits. While we did not explore this possibility, it represents a potentially important effect that warrants future investigation.

An additional caveat is that we define radial redistribution through fractional changes in $j_\phi$, thereby focusing on stars that redistributed by being torqued. If instead we adopted other definitions of radial redistribution \citep[as in][]{Bellardini2026}, then we would find even lower cold-torqued fractions among redistributed stars. In other words, our results represent an upper limit to how `cold' radial redistribution is. We show this in Appendix~\ref{sec:rr_metrics} where we compare changes in eccentricity across redistribution metrics $j_\phi$, $R(j_\phi)$, and $R$.

These FIRE-2 simulations do not include all relevant physical processes that act in galaxies. Some of the physics not included in these simulations, such as magnetohydrodynamics (MHD) and cosmic-ray feedback, is unlikely significantly to change our conclusions. Although cosmic-ray feedback can have significant implications on galaxy-wide properties \citep{Hopkins2020b, Chan2022}, \cite{McCluskey2024} showed that cosmic-ray feedback has little impact on the dynamical state of stellar disks, in terms of $v_{\phi} /  \sigma_{v}$, at fixed stellar mass.

In particular, the absence of supermassive black holes and AGN feedback could affect our results. AGN feedback is expected to regulate the gas content and star formation of galaxies, especially in their central regions \citep{SilkRees1998, Fabian2012, HeckmanBest2014, Alexander2025}, and in its absence galaxies can remain more centrally concentrated and sustain enhanced central star formation, resulting in greater velocity dispersions and stronger outflows \citep{Chan2022, Gandhi2022, Wellons2023, Marasco2025, Liu2026}. The lack of AGN may lead these FIRE-2 galaxies to be more centrally concentrated, and this may affect the prevalence of cold-torqued stars in the inner regions.

Additionally, FIRE-2 galaxies rarely host strong, long-lived bars \citep{Ansar2025}. This is important for our analysis, because bars can influence stellar orbits in two distinct ways. On the one hand, bars can act as non-axisymmetric heating agents that increase orbital eccentricities and contribute to hot redistribution, with simulations finding that stronger bars are associated with stronger heating, particularly for young stars \citep{Grand2016a}. On the other hand, bars can also drive coherent changes in angular momentum, including outward migration associated with bar formation, bar slowdown, and resonant trapping \citep{Chiba2021, Baba2025}. Some works argue that bar-driven redistribution is a major contributor to cold torquing in the MW \citep{Haywood2024, Zhang2025bar}. Consistent with this, \cite{Okalidis2022} found in the Auriga simulations that strongly barred galaxies show enhanced redistribution relative to weakly barred or non-barred systems, although they did not disentangle between cold and hot torquing. As a result, our simulations may underestimate the prevalence of stars being cold torqued in the MW. At the same time, because bars also can heat stellar orbits, the lack of strong bars in FIRE-2 also may lead to less hot redistribution. It is therefore not obvious whether weaker bars cause us to overestimate or underestimate the cold-torqued fraction overall, although they likely reduce the contribution of cold torquing from bars relative to strongly barred systems. Using the bar strengths that \cite{Ansar2025} measured for the FIRE-2 MW-mass galaxies, we do not find that the galaxies with the strongest bars have unusually high or low amounts of radial redistribution or its coldness, although a more detailed comparison with bar strength and other possible drivers is beyond the scope of this work.

Finally, these FIRE simulations represent a cosmologically representative sample of MW/M31–mass galaxies in isolated or Local Group–like environments, and therefore describe general behavior of MW/M31-mass galaxies rather than the specific evolutionary history of the MW. 
Stars in FIRE-2 simulations generally have hotter kinematics than those observed in the MW for stars older than a few 100 Myr \citep{McCluskey2025}. This is important for interpreting our results, because we find that the cold-torqued fraction is highest in earlier-forming and dynamically colder disks. This suggests that stars in FIRE-2 simulations may experience more radial redistribution from dynamical heating than in the MW. Consequently, our results may underestimate the prevalence of cold-torqued stars relative to what would occur in a dynamically-colder and earlier-forming disk such as the MW, at least for young stars. However, \cite{McCluskey2025} showed that the MW itself is a kinematic outlier compared with M31, M33, and several PHANGS galaxies, and that these FIRE simulations agree reasonably well with them.

For this reason, we also examined Romeo separately. Romeo is one of our earliest-forming disks, and it formed in a Local Group–like environment, so it likely serves as our best MW analog, and among our FIRE-2 sample, it has $v_\phi / \sigma_{v, \rm 3D}$ most similar to the MW. Even so, the trends in radial redistribution in Romeo are broadly consistent with those found across the full FIRE-2 sample.
Thus, our analysis of Romeo suggests that any differences for the MW are likely by degree, and that the general trends should hold.

\subsection{Connections to Other Works}

Many works discuss radial redistribution/migration in the context of corotation-resonance scattering, in which stars exchange angular momentum with non-axisymmetric structures while remaining on near-circular orbits \citep{SellwoodBinney2002}.
Although some stars are cold torqued between birth and today in the FIRE cosmological simulations, radial redistribution more commonly includes dynamical heating, except for the youngest stars that redistributed outwards.

Many works have examined cold torquing in idealized analytic or $N$-body disks \citep[for example][]{Roskar2008b, Roskar2012, Grand2015a, Halle2015, Halle2018, Haywood2024} and few works have examined radial redistribution more generally in cosmological simulations \citep[for example][]{Grand2016a, verma2021, Lu2022, Vincenzo2020, Boecker2022, Okalidis2022}.
However, few works explicitly tested how often significant redistribution preserves eccentricity in cosmological disks, especially without restricting the analysis to stars born on circular orbits.

Several works showed that, in realistic disks with complex non-axisymmetric structures and overlapping resonances (such as bar-spiral interactions), redistribution is often not dynamically cold and can be accompanied by changes in orbital actions \citep[for example][]{MinchevFamaey2010, Minchev2011, Minchev2012a, Daniel2019}. \cite{zhang2025_highz} similarly found that gas-rich disks can drive more efficient radial redistribution while also increasing the heating-to-migration ratio. Our results are broadly consistent with that picture, in the sense that in our simulations radial redistribution is typically not cold. However, we do not directly identify the dynamical drivers of the redistribution or heating. The FIRE simulations contain many of the structures commonly associated with heating as found in the studies mentioned above, including spiral structure \citep{Orr2023, Quinn2025}, bars \citep{Ansar2025}, and giant molecular clouds \citep{Benincasa2020, Guszejnov2020}, but determining which of these processes drive heating or cooling is beyond the scope of this work.

The strongest trend we identify is the asymmetry between inward and outward migrators. \cite{Halle2015} found that outward migrators tend to decrease their eccentricities, while inward migrators tend to increase them. However, they interpreted this with caution, because of how they measured eccentricity. Our results show a qualitatively similar trend using a more robust measurement of eccentricity. Outward migration is the regime in which being cold(er)-torqued is most common, whereas inward migration almost always leads to dynamical heating. The results of \cite{Bird2012} support this directional dependence: they showed that changes in circularity are tied to whether a star moves inward or outward, particularly in the presence of satellite perturbations. \cite{Minchev2012a} discussed a qualitatively similar asymmetry and interpreted it through the conservation of radial action, such that stars moving inward into a deeper potential tend to heat while stars moving outward tend to cool. That interpretation provides one possible explanation that is qualitatively consistent with our results, but it also relies on assumptions about the nature of the perturbations and the orbital evolution along the way, which we do not test here.

Our identification of a substantial population of stars that dynamically cooled in strong association with outward redistribution suggests that radial redistribution should not be framed only in terms of whether it is cold or hot relative to birth.
\cite{GonzalezRiveradeLaVernhe2024} identified a population of extremely metal-poor stars in the MW on cold orbits.
Previous works have shown that the cosmological formation histories of MW-mass galaxies generically lead to arbitrarily metal-poor stars being on orbits prograde with the disk today \citep{Santistevan2021, Sestito21}, but it is unclear if this can explain the fraction of such stars on cold, near-circular orbits today.
Dynamical processes, such as via the Inner Lindblad Resonance, can change $j_R$,
and \cite{Smock2026} showed how this can lead to dynamical cooling of orbits.
Other works have proposed dynamical cooling in the vertical direction from changes in $j_\phi$ at corotation \citep{Minchev2012b, VeraCiro2016}, though they did not explore the link to changes in $j_R$.
\cite{Khoperskov2020}, using an $N$-body simulation of a MW-like galaxy, found that stars from the inner disk can redistribute outward and eventually settle onto nearly circular orbits; our analysis is consistent with the existence of such a population.
We showed that the outcomes of radial redistribution are not limited to preserving or erasing circularity, but also can include substantial cooling of orbits.
Finding stars on circular orbits today does not imply that they preserved their circularity throughout radial redistribution.

Many works characterized radially redistributed stars as a biased subset of the disk, such that cold torquing preferentially occurs for dynamically cold stars \citep{Solway2012, VeraCiro2014, Halle2015, Halle2018, Daniel2018, Mikkola2020}.
Although the literature often suggests that stars on the coldest and most circular orbits should be the most susceptible to specific migration mechanisms, we do not find that stars born on the most circular orbits are the stars most likely to preserve their eccentricity. Instead, among stars that underwent significant radial redistribution, those born on moderately eccentric orbits are the most likely to show little net change in eccentricity. This is not necessarily inconsistent with previous works, because they examined how efficiently a specific mechanism operates for stars of different orbital properties. Our analysis instead asks, given that a star redistributed, how likely was it to preserve its orbital shape? In that sense, preserving eccentricity alone does not uniquely identify classical cold torquing through corotation resonances. Some of the stars that we classify as cold torqued instead may represent a population that experienced multiple heating and cooling events, leaving little net change in eccentricity overall \citep{Struck2026}.

We also find that cold-torqued stars are more prevalent in earlier-forming and dynamically colder disks. This relates to previous works that showed that dynamically colder disks host stronger spiral arms \citep{Bird2012, Quinn2025, Ghosh2025}, which can be efficient drivers of cold torquing \citep[for example][]{SellwoodBinney2002, Roskar2008b, Roskar2012, Daniel2015}. This provides a plausible interpretation for why colder disks in our sample exhibit higher cold-torqued fractions, although we do not directly test whether the spiral arms themselves are the causal driver of that trend. In light of these results, when thinking about radial redistribution in the context of stellar eccentricities, one must distinguish between the dynamical state of the disk as a whole and the birth eccentricity of an individual star. Our analysis shows that preserving eccentricity does not always follow the classic cold-torquing scenario in \cite{SellwoodBinney2002}, and that even in dynamically cold disks, the outcome depends strongly on stellar age and migration direction.

Observational studies of the MW also have argued that cold torquing is an important mode of radial redistribution \citep[for example][]{Frankel2020, Feltzing2020, Lian2022, Lehmann2024}. However, these studies necessarily rely on assumptions of the metallicity and dynamical history of the disk and selection choices. For example, \cite{Frankel2020} modeled coeval stars as being born on near-circular orbits, while \cite{Feltzing2020} found a cold-torqued fraction of 10\%, requiring such stars to reside on highly circular orbits today. Additionally, \cite{Lehmann2024} assumed that stars on circular orbits today either have not left their birth orbit or have been churned. Such assumptions in some cases preferentially may isolate stars that are circular today, including stars that dynamically cooled during outward migration, rather than uniquely identifying stars that preserved circularity throughout their radial redistribution. We therefore view these MW studies as broadly consistent with the idea that radial redistribution is important, while also emphasizing that observational claims about the prevalence of cold torquing depend sensitively on how one treats stellar birth conditions and present-day orbital circularity.

Finally, we reemphasize that we only examined changes in orbital eccentricity between a star particle's birth and today. Thus, the relatively low cold-torqued fractions that we measure do not automatically imply that cold torquing rarely occurs, but that if it does, other dynamical processes that also affect eccentricities often occur between birth and today.
In dynamically active, cosmological disks, the same star can be affected by multiple processes over time, such that cold torquing, heating, and cooling all can contribute to its present-day orbit. Therefore, a star that does not preserve its eccentricity since birth has not necessarily failed to undergo cold torquing during its lifetime, and a star that is on a near-circular orbit today has not necessarily preserved circularity throughout its radial redistribution. Our results instead show that a star's present-day orbit reflects the cumulative outcome of a star's migration direction, birth eccentricity, age, and subsequent dynamical history. Therefore, our results point to the difficulty in inferring the degree of hot or cold torquing from present-day eccentricity alone.

\section*{Acknowledgements}

We performed this work using the \textsc{GizmoAnalysis} package \citep[][]{Hopkins2015}, the Astropy package \citep[][]{Astropy2013, Astropy2018}, as well as libraries from Numpy \citep[][]{Numpy2020}, SciPy \citep[][]{Scipy2020}, and Matplotlib \citep[][]{Matplotlib2007}.

CS and FM received support from NASA FINESST award 80NSSC24K1484.
AW received support from: NSF via CAREER award AST-2045928.

We generated simulations using: XSEDE, supported by NSF grant ACI-1548562; Blue Waters, supported by the NSF; Frontera allocations AST21010 and AST20016, supported by the NSF and TACC; Pleiades, via the NASA HEC program through the NAS Division at Ames Research Center.

The FIRE-2 simulations are publicly available \citep{Wetzel2023, Wetzel2025} at \url{http://flathub.flatironinstitute.org/fire}.
Additional FIRE simulation data is available at \url{https://fire.northwestern.edu/data}.
A public version of the GIZMO code is available at \url{http://www.tapir.caltech.edu/~phopkins/Site/GIZMO.html}.

\bibliographystyle{mnras}
\bibliography{biblio}{}

\appendix

 \section{A. Correlation with Disk Onset and $v_{\phi}/\sigma_{v, \rm 3D}$}

\renewcommand{\thetable}{A\arabic{table}}
\setcounter{table}{0}
\label{sec:corr_coeff}

\begin{table*}[!t]
\centering
\caption{
Correlation coefficients between the cold-torqued fraction and either the disk onset times or the disk's $v_{\phi} / \sigma_{v, \rm 3D}$ for young stars today.
}
\label{tab:ct_correlations}

\centering
\setlength{\tabcolsep}{3pt}
\renewcommand{\arraystretch}{1.15}

\resizebox{\linewidth}{!}{%

\begin{tabular}{|@{\hspace{8pt}}l@{\hspace{4pt}}|c|c|c|c||c|c|c|c||c|c|c|c|}
\hline
& \multicolumn{4}{c||}{early-disk onset time}
& \multicolumn{4}{c||}{late-disk onset time}
& \multicolumn{4}{c|}{$v_{\phi}/\sigma_{v,\mathrm{3D}}$ of young stars} \\
\hline

Sample
& Pearson $r$ & $p$
& Spearman $\rho$ & $p$
& Pearson $r$ & $p$
& Spearman $\rho$ & $p$
& Pearson $r$ & $p$
& Spearman $\rho$ & $p$ \\
\hline

\textit{age $<2\,\mathrm{Gyr}$}
& {} & {} & {} & {}
& {} & {} & {} & {}
& {} & {} & {} & {} \\

outward migrators
& 0.87 & $2.7\times10^{-4}$
& 0.88 & $2.6\times10^{-4}$
& 0.55 & 0.06
& 0.60 & 0.04
& 0.94 & $7.4\times10^{-6}$
& 0.89 & $1.1\times10^{-4}$ \\

inward migrators
& 0.70 & 0.01
& 0.76 & $4.3\times10^{-3}$
& 0.58 & 0.05
& 0.66 & 0.02
& 0.80 & $2.0\times10^{-3}$
& 0.85 & $4.2\times10^{-4}$ \\
\hline

\textit{born during late-disk}
& {} & {} & {} & {}
& {} & {} & {} & {}
& {} & {} & {} & {} \\

outward migrators
& 0.71 & 0.01
& 0.66 & 0.02
& 0.24 & 0.44
& 0.31 & 0.32
& 0.95 & $3.6\times10^{-6}$
& 0.93 & $1.2\times10^{-5}$ \\

inward migrators
& 0.37 & 0.24
& 0.50 & 0.11
& 0.28 & 0.38
& 0.27 & 0.40
& 0.67 & 0.02
& 0.77 & $3.5\times10^{-3}$ \\
\hline

\end{tabular}%
}
\tablecomments{
We list Pearson $r$ and Spearman $\rho$ with their respective $p$-values.
We show results separately for outward and inward migrators, and for stars either $<2 \Gyr$ old or all stars born during the late-disk era.
}
\end{table*}

Table~\ref{tab:ct_correlations} presents the correlation coefficients for the trends in Figure~\ref{ct_onset_vsig}, which we measure separately for stars younger than $2 \Gyr$ today, as well as for all stars born during the late-disk era. For the late disk sample, we use the mean $v_{\phi} / \sigma_{v, \rm 3D}$ of all late-disk stars, whereas in Figure~\ref{ct_onset_vsig}, we calculate $v_{\phi} / \sigma_{v, \rm 3D}$ using stars younger than $100 \Myr$.

\section{B. Cold-Torqued Thresholds}
\label{sec:ct_thresh}

\renewcommand{\thefigure}{B\arabic{figure}}
\setcounter{figure}{0}

\begin{figure}[!ht]

\includegraphics[width = 0.975 \linewidth]{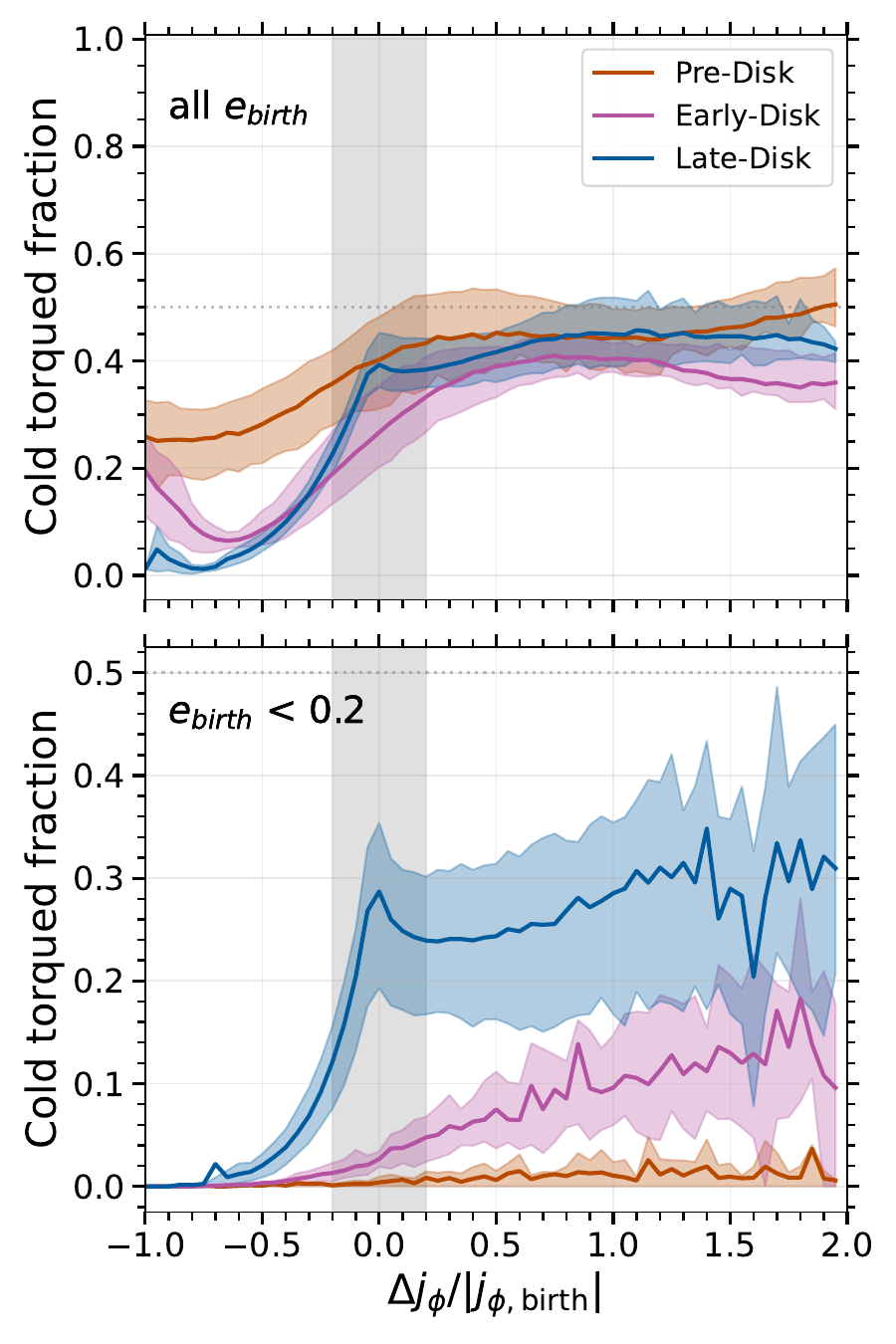}
\caption{
Cold-torqued fraction versus $\Delta j_{\phi} / |j_{\phi,\rm birth}|$ for all stars regardless of birth eccentricity (top) and stars born on circular orbits ($e_{\rm birth} < 0.2$; bottom), 
for stars born in each dynamical era.
Lines show the mean, shaded regions show the 68th percentile scatter across the 12 galaxies. Late-disk stars born on circular orbits can experience large fractional changes in angular momentum while maintaining a relatively constant cold-torqued fraction.
}
\label{ct_vs_delj}
\end{figure}

\begin{figure}[!ht]

\includegraphics[width = 0.975 \linewidth]{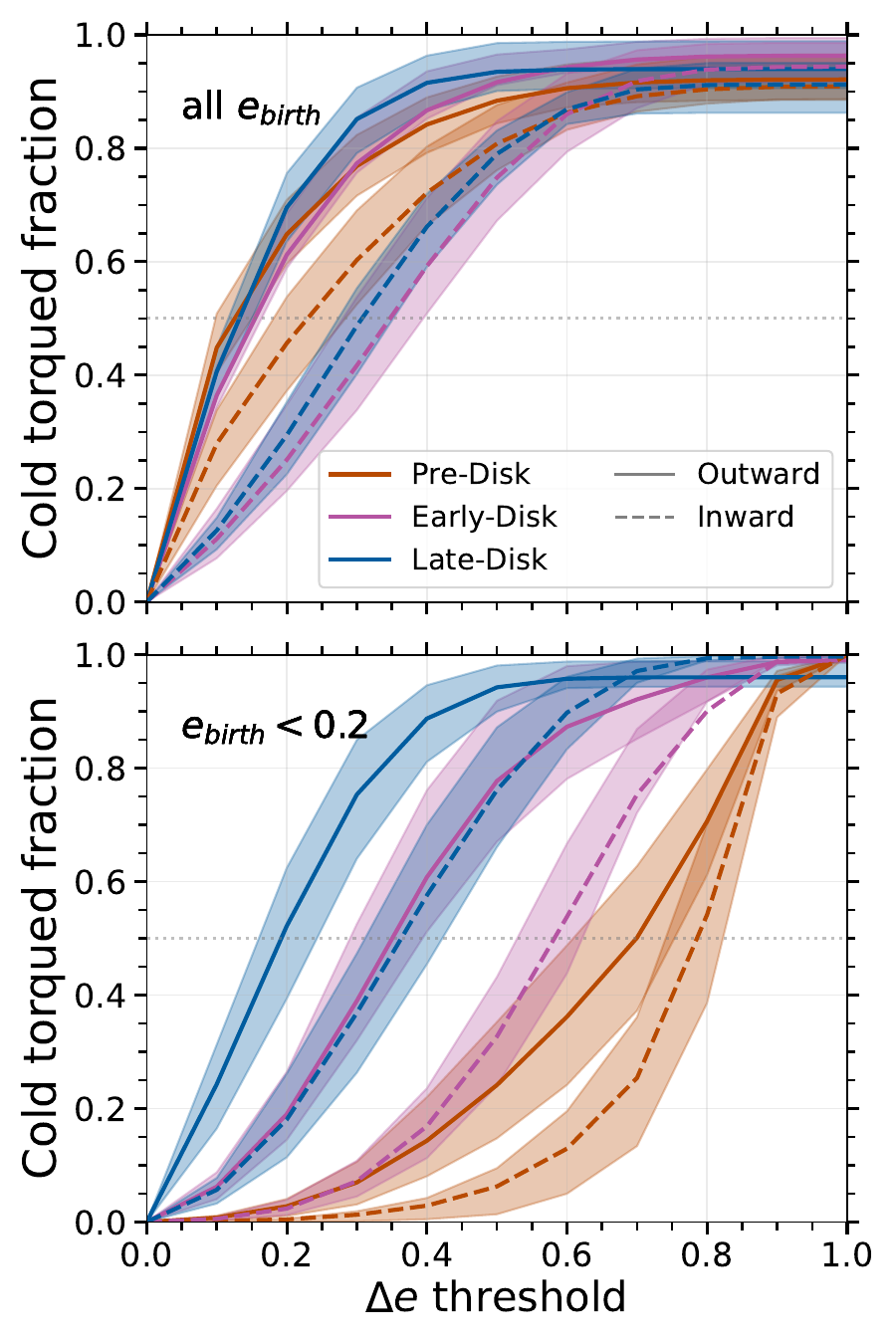}
\caption{
Cold-torqued fraction versus the maximum $|\Delta e|$ for all stars (top) and stars born on near-circular orbits ($e_{\rm birth} < 0.2$; bottom), for stars born in the three dynamical eras, separated by migration direction. Lines show the mean, shaded regions show the 68th percentile scatter across the 12 galaxies.
The cold-torqued fractions across the three dynamical eras remain robust and consistent regardless of the specific thresholds used to define `cold' dynamical evolution.
}
\label{ct_vs_dele_thresh}
\end{figure}

Figure~\ref{ct_vs_delj} shows the cold-torqued fraction versus $\Delta j_{\phi} / |j_{\phi,\rm birth}|$ for all stars (left) and for stars born on nearly circular orbits ($e_{\rm birth} < 0.2$; right), separately for stars born in each of the 3 eras of disk evolution. Across both $e_{\rm birth}$ thresholds, the cold-torqued fraction is approximately flat at $\Delta j_{\phi} / |j_{\phi,\mathrm{birth}}| > 0$, indicating that the amount of radial redistribution toward larger radii has little effect on whether stars migrate dynamically cold. In contrast, for inward migration ($-1 < \Delta j_{\phi} / |j_{\phi,\mathrm{birth}}| < 0$) and without imposing a birth eccentricity cutoff, the cold-torqued fraction reaches a minimum, with late-disk stars approaching 0\%, early-disk stars $\approx$ 10\%, and pre-disk stars $\approx$ 25\%. For $\Delta j_{\phi} / |j_{\phi,\mathrm{birth}}| < -1$, which we also classify as outward migration, the cold-torqued fraction increases again. In the right panel, which is only for stars born on near-circular orbits, the cold-torqued fraction instead declines sharply with inward migration and is zero for $\Delta j_{\phi} / |j_{\phi,\mathrm{birth}}| \lesssim -0.5$.

We replicated this figure for stars that span a variety of birth radii and present-day radii, and we also verified that the result is insensitive to the specific migration metric adopted. When we quantify radial redistribution using changes in guiding-center radius, rather than fractional changes in angular momentum, we recover the same qualitative behavior: a nearly flat trend for outward migrators and a rapidly declining trend for inward migrators. Thus, the sharp asymmetry between inward and outward migration is not caused by a particular choice of sample selection or threshold in radial redistribution.

Figure~\ref{ct_vs_dele_thresh} shows the cold-torqued fraction versus the maximum allowed change in eccentricity, $|\Delta e|$, used to define `cold' for inward and outward migrators. We further divide the sample according to birth eccentricity: the top panels include all stars, while the bottom panels include only stars formed on near-circular orbits ($e_{\mathrm{birth}} < 0.2$). 
The cold-torqued fraction increases monotonically with the allowed change eccentricity across all three disk eras and migration direction. Relaxing the definition of `cold' naturally includes a larger fraction of stars. However, an important distinction emerges between the top (all $e_{\mathrm{birth}}$) and bottom ($e_{\mathrm{birth}} < 0.2$) panels for both thresholds. When we include all stars, the behavior across disk eras appears more self-similar, reflecting the fact that stars born on hotter orbits are more likely to remain comparably `cold' relative to their initial eccentricities. In contrast, when restricting to stars born on near-circular orbits, the differences between disk eras become more pronounced but remain different across the varied thresholds.

We also varied the cold-torqued fraction with the minimum threshold in $|\Delta j_{\phi} / j_{\phi,\mathrm{birth}}|$ to define `significant' radial redistribution and found the trend to be relatively flat across all three disk eras and both inward and outward migrators. Therefore, our results are not strongly sensitive to the choice of redistribution threshold, and varying this threshold does not alter our key conclusions.

\section{C. Metrics of Radial Redistribution}

\renewcommand{\thefigure}{C\arabic{figure}}
\setcounter{figure}{0}
\label{sec:rr_metrics}

\begin{figure*}[!ht]
\centering
\includegraphics[width = 0.975 \linewidth]{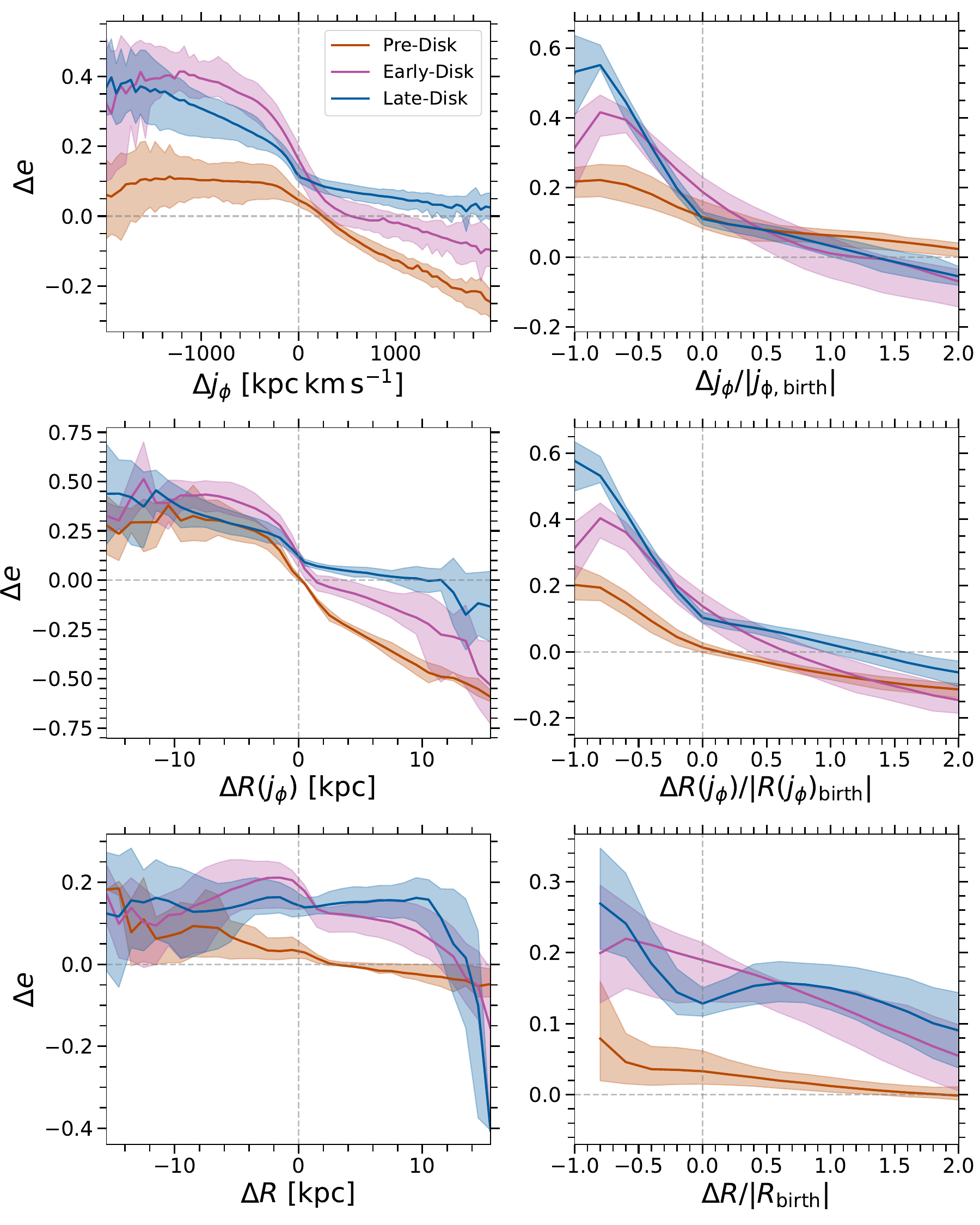}
\caption{Median change in eccentricity, $\Delta e$, as a function of six different radial redistribution metrics for the three disk eras. The panels compare fractional and absolute changes in specific angular momentum, $j_\phi$, the radius of a circular orbit with the same angular momentum ( guiding-center radius), $R(j_\phi)$, and the instantaneous galactocentric radius, $R$. Lines represent the mean and shaded regions indicate the 68th percentile scatter across the 12 galaxies. Fractional changes in $j_\phi$ and $R(j_\phi)$ provide consistent metrics for diagnosing radial redistribution, whereas $R$ is difficult to use for identifying cold-torqued stars, because it is inherently sensitive to eccentricity.
\label{dele_rs}}
\end{figure*}

Figure~\ref{dele_rs} shows the median change in eccentricity, $\Delta e$, for all three eras as a function of six metrics of radial redistribution, considering both absolute and fractional changes in $R$ (instantaneous galactocentric radius), $R_{\rm circ}(j)$ (guiding-center radius), and $j_{\mathrm{\phi}}$ (angular momentum). The fractional changes in $j_{\mathrm{\phi}}$ and $R_{\rm circ}(j)$ are consistent, and $\Delta j$ and $\Delta R_{\rm circ}(j)$ also agree. In contrast, the two metrics that use $R$ generally produce positive changes with $\Delta e$, demonstrating how sensitive changes to $R$ are to eccentricity. This occurs because instantaneous $R$ depends on orbital phase as well as the underlying orbit. Thus, stars on eccentric orbits can have large changes in $R$ between birth and today even if their guiding centers or angular momenta have not changed substantially.

\section{D. Radial Dependence of Eccentricity}
\label{sec:radial_e}

\renewcommand{\thefigure}{D\arabic{figure}}
\setcounter{figure}{0}

\begin{figure*}[!ht]
\centering
\includegraphics[width = 0.975 \linewidth]{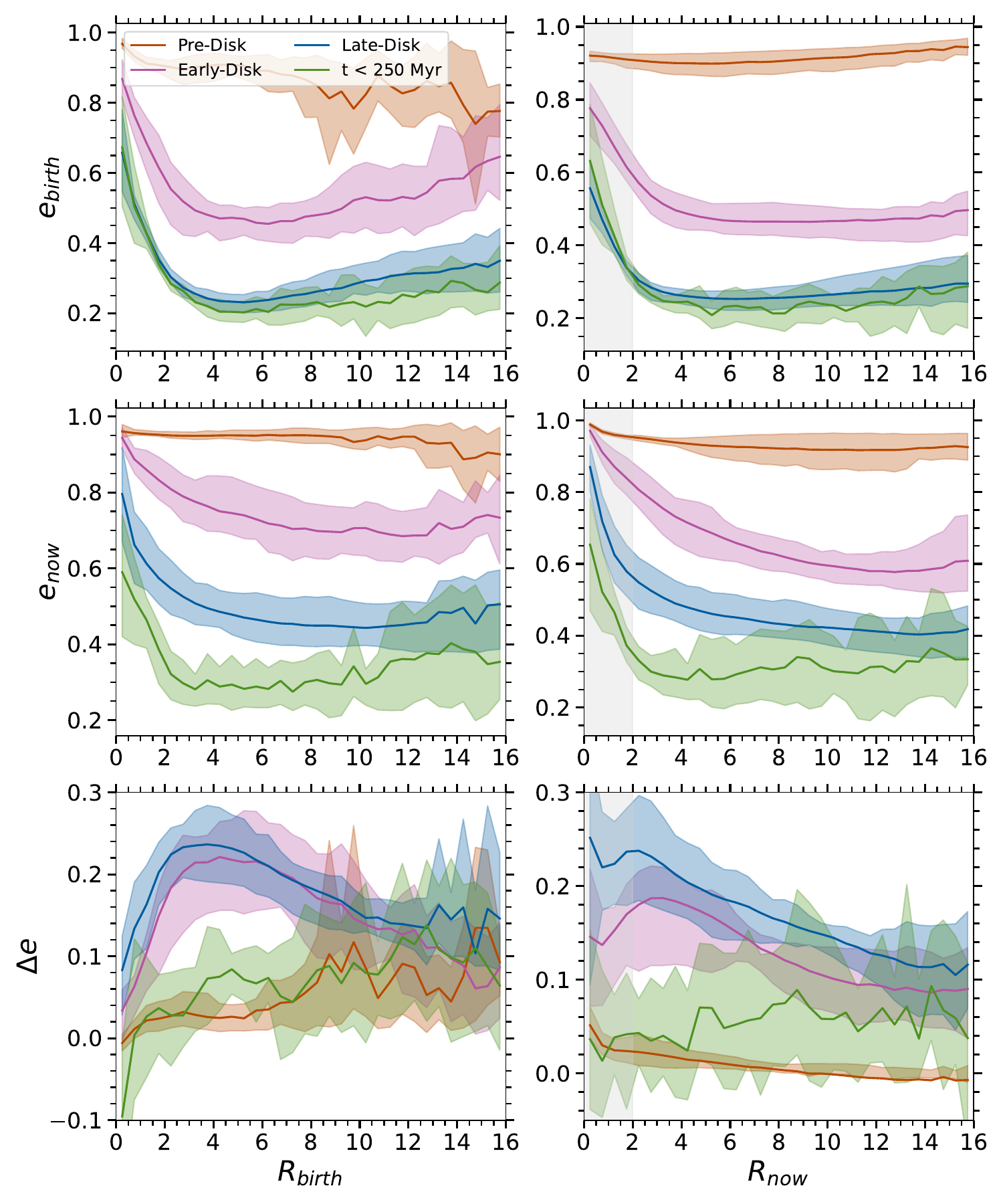}
\caption{
Radial dependence of stellar eccentricity and its evolution from birth to today. The left column shows quantities versus birth radius, $R_{\rm birth}$, while the right column shows the same quantities versus present-day radius, $R_{\rm now}$. The top row shows birth eccentricity, $e_{\rm birth}$, the middle row shows present-day eccentricity, $e_{\rm now}$, and the bottom row shows the change in eccentricity, $\Delta e$. Lines show stars born in each of the three disk eras, and the green line shows stars younger than 250 Myr. Lines show the mean and shaded regions show the 68th percentile scatter across the 12 galaxies. Stars born in the inner galaxy were on more eccentric orbits and experienced a smaller net change in eccentricity than stars born maximally circular at larger radii.
}
\label{efen_rs}
\end{figure*}

Figure~\ref{efen_rs} shows the radial dependence of stellar eccentricities and their evolution. The left column shows quantities as a function of birth radius, $R_{\rm birth}$, while the right column shows the same quantities as a function of present-day radius, $R_{\rm now}$. The top row shows the birth eccentricity, $e_{\rm birth}$, the middle row shows the present-day eccentricity, $e_{\rm now}$, and the bottom row shows the change in eccentricity, $\Delta e$. Lines are shown separately for the three disk eras, with the green line indicating stars younger than 250 Myr. We show the radial dependence of $\Delta e$ to provide additional context for interpreting our results.

In the pre-disk era, $\Delta e$ is largely flat as a function of radius, consistent with the absence of a dynamically settled disk. In the early- and late-disk eras, $\Delta e$ peaks at the radius where stars are born maximally circular, a result that is physically intuitive and reflects the role of dynamical birth conditions, in that stars born on more circular orbits are more susceptible to subsequent dynamical heating. Toward smaller radii, $\Delta e$ declines despite the presence of strong perturbations such as a bar, because stars are born on intrinsically hotter orbits, which limits further changes in eccentricity rather than indicating reduced heating efficiency. This produces a clear downturn at small radii that is driven by birth conditions. The green line shows these trends for stars younger than 250 Myr, exhibiting largely flat radial trends, suggesting that these stars have not yet experienced sufficient dynamical evolution simply because they have not had time to be heated. At larger radii, the decrease in $\Delta e$ is consistent with lower surface densities, implying that where stars end up relative to where they were born plays an important role in determining whether they heat or cool, and tying back to the differences observed between inward and outward migrators.

\end{document}